\documentclass[a4paper, 11pt]{article}
\usepackage{amsmath}
\usepackage{amssymb}
\usepackage{setspace}
\usepackage{xcolor}
\usepackage{graphicx}
\usepackage{placeins}
\usepackage[english]{babel}
\usepackage{apacite}
\usepackage{caption}
\usepackage{booktabs}
\usepackage{multirow}
\usepackage{minibox}
\usepackage{authblk}
\usepackage{epstopdf}
\usepackage[hmargin = 2cm, vmargin = 2.5cm]{geometry}
\usepackage{natbib}

\newcommand{\E}{\mathop{\mathbb{E}}}

\newcommand{\erf}{\mathop{\mathrm{erf}}}

\newcommand{\ddd}{,\,\ldots,\,}

\newtheorem{theorem}{Theorem}[section]
\newtheorem{definition}{Definition}[section]
\newtheorem{proposition}{Proposition}[section]

\newtheorem{remark}{Remark}[section]

\newtheorem{proof}{Proof}[section]

\begin{document}

\title{Estimating Quantile Families of Loss Distributions for Non-Life Insurance Modelling via L-moments}
\author[1,2,3]{Gareth W. Peters}
\author[4]{Wilson Y. Chen}
\author[4]{Richard H. Gerlach}
\affil[1]{Department of Statistical Science, University College London, UK}
\affil[2]{Oxford-Man Institute, Oxford University, UK}
\affil[3]{System Risk Center, London School of Economics, UK}
\affil[4]{Discipline of Business Analytics, University of Sydney}
\date{February, 2016}
\maketitle

\begin{abstract}
This paper discusses different classes of loss models in non-life insurance settings. It then overviews the class Tukey transform loss models that have not yet been widely considered in non-life insurance modelling, but offer opportunities to produce flexible skewness and kurtosis features often required in loss modelling. In addition, these loss models admit explicit quantile specifications which make them directly relevant for quantile based risk measure calculations. We detail various parameterizations and sub-families of the Tukey transform based models, such as the g-and-h, g-and-k and g-and-j models, including their properties of relevance to loss modelling.

One of the challenges with such models is to perform robust estimation for the loss model parameters that will be amenable to practitioners when fitting such models. In this paper we develop a novel, efficient and robust estimation procedure for estimation of model parameters in this family Tukey transform models, based on L-moments. It is shown to be more robust and efficient than current state of the art methods of estimation for such families of loss models and is simple to implement for practical purposes.
\end{abstract}

\section{Introduction: context of modelling losses in general insurance}

In general, one can consider insurance to be principally a data-driven industry in which the primary cash out-flows comprise of claim payments. Insurance companies therefore employ large numbers of analysts which include actuaries, to understand the claims data. There are many categories of insurance business lines from which claims payment flows arise. The types of insurance business lines considered in this manuscript will be the area of general insurance and non-life insurance, which is probably one of the most active areas of research in actuarial science. General insurance is typically defined as any insurance that is not determined to be life insurance. It is called property and casualty insurance in the U.S. and Canada and Non-Life Insurance in Continental Europe. Non-Life insurance typically includes modelling lines of business such as health insurance, personal/property insurance such as home and motor insurance as well as large commercial risks and liability insurance. 

When considering the claim payments in non-life lines of business, traditionally the claim actuaries are concerned with the amount the insurance company will have to pay. Therefore, the general insurance actuary needs to have an understanding of the various models for the risk consisting of the total or aggregate amount of claims payable by an insurance company over a fixed period of time. Insurance data that they may consider fitting statistical models to contains relatively large claim amounts, which may be infrequent. Therefore, an important aspect of actuarial science involves the development of statistical models that can be utilised to describe the claims process accurately so that reserves and liability management can be accurately performed. The models such non-life insurance actuaries develop should enable them to accurately make decisions on things such as: premium loading, expected profits, reserves necessary to ensure (with high probability) profitability and the impact of reinsurance and deductibles. In particular, a core role for a non-life insurance actuary involves preserving the insurance company’s financial security by accurately estimating future claims liabilities. Reserving for the amount of future claims payments involves a large degree of uncertainty, especially for long tail class business where tail behaviors can be largely different. Hence, it can be difficult to estimate the loss reserve precisely. Fortunately, there are now numerous classes of models that have been developed for modelling claims in non-life insurance settings, for excellent reviews see \cite{wuthrich2008stochastic}, \cite{klugman2012loss}, \cite{denuit2006actuarial}, \cite{mcneil2015quantitative} and models discussed in a similar context for heavy tailed, lepto-kurtic and plato-kurtic loss models in \cite{peters2015advances}. 

In particular, many studies have pointed to the importance of flexible models with a variety of skew and kurtosis properties to be considered in modelling non-life insurance loss processes. In practice, it is popular too consider two parameter shape and scale models such as log-normal, gamma, Weibull and Pareto models, see \cite{taylor2012loss} for discussions. Primarily, these models have been popular due to simplicity in parameter estimation and model selection. However, it has been observed that making these distributional assumptions, actuaries may underestimate the risk inherited in the long tail which is affected by large claim liabilities because these distributions do not possess flexible tails to describe the features of large claims. Failure to estimate the large claim liabilities adequately can cause financial instability of the company and eventually lead to insolvency. In order to improve modeling accuracy and reliability, a number of more sophisticated models have been studied for such loss modelling. These include the Poisson-Tweedie family of models in the additive exponential dispersion class, see discussion in \cite{peters2009model}, the GB2 models studied in \cite{cummins1990applications} and \cite{dong2013bayesian}, the generalized-t (GT) \cite{chan2008robust}, the Stable family \cite{paulson1985practical} and \cite{peters2011analytic} and the Pearson family \cite{aiuppa1988evaluation}. In this manuscript, we aim to raise awareness in the actuarial community of another alternative class of models that can be considered for such loss modelling based on different variations of the Tukey family.

We note that, in practice, many practitioners are reluctant to utilise such flexible models due to complications that can arise in real applications relating to ease of parameter estimation and model selection. These can play important practical limitations for utilisation of such models. In this paper, we will therefore focus on two aspects, first the introduction of an under utilised family of flexible skew and kurtosis models for such non-life insurance modelling applications and then secondly a novel accurate and robust estimation procedure that we have developed to fit such models in practice. This will allow the general Tukey transform models such as g-and-h, g-and-k and g-and-j models considered in this paper to be easily implemented in practice. We show that our proposed estimation method is robust and more accurate than all current competing methods. The novel approach we develop is based around L-moments based estimation. We will discuss in detail the family of models we present, introducing the actuarial literature general forms of the Tukey families follows by developing our estimation procedure before presenting applications of the model and the new estimation approach.

\section{General Families of Quantile Transform Distributions For Insurance Modelling}
Here we discuss several distributional families relevant to loss modeling in insurance which can only be specified via the transformation of another standard random variable, for example a Gaussian. Examples of such models which are typically defined through their quantile functions include the Johnson family with base distribution given by Gaussian or logisitic and the Tukey family with base distribution typically given by a Gaussian or logistic. The concept of constructing skew and heavy-tailed distributions through the use of a transformation of a Gaussian random variable was originally proposed in the work of \cite{Tukey1977} and is therefore aptly named the family of Tukey distributions. This family of distributions was then extended by \cite{Hoaglin1985}, \cite{jorge1984some}, \cite{azzalini1985class} and \cite{fischer2007tukey}. The multivariate versions of these models have been discussed by \cite{field2012multivariate}.

Within this family of distributions, two particular subfamilies have received the most attention in the literature; these correspond to the g-and-h and the g-and-k distributions. The first of these families the g-and-h has been studied in a several contexts, see for instance the developments in the areas of risk and insurance modelling in \cite{dutta2006tale}, \cite{peters2006bayesian}, \cite{degen2007quantitative}, \cite{jimenez2011using} and the detailed discussion in \cite[Chapter 9]{cruz2015fundamental}.  The second family of g-and-k model has been looked at in works such as \cite{Haynes1997} and \cite{hossain2009numerical}.

The advantage of models such as the g-and-h family for modeling losses in a non-life insurance setting is the fact that they provide a very flexible range of skew, kurtosis, and heavy-tailed features while also being specified as a rather simple transformation of standard Gaussian random variates, making simulation under such a models efficient and simple. It is important to note that the support of the g-and-h density includes the entire real line, in some cases this is appropriate in non-life insurance modelling such as under logarithmic transfoms of the loss or claim amounts. In other cases it may be more appropriate to consider loss models with strictly positive supports, in general this can be achieved either by truncation or by restriction of the parameter values. In some subfamily members, the g-and-h family automatically takes a positive support such as the Double h--h subfamily. 

\subsection{Tukey's Elongation Transform Family of Loss Models}
We begin by discussing the general family of Tukey distributions. Basically, Tukey suggested several nonlinear transformations of a standard Gaussian random variable, denoted below throughout by $W \sim \mathrm{Normal}(0,1)$. The g-and-h transformations involve a skewness transformation $g$ and a kurtosis transformation $h$. If one replaces the kurtosis transformation of the type $h$ with the type $k$, one obtains the g-and-k family of distributions discussed by \cite{RaynerMacGillivray2002}. If the $h$ transformation is replaced by the $j$ transformation, one obtains the g-and-j transformations of \cite{fischer2004kurtosis}.

The generic specification of the Tukey transformation is provided in Definition~\ref{DefnTukey}. These types of transformations were labelled elongation transformations, where the notion of elongation was noted to be closely related to tail properties such as heavy-tailedness (see discussions by \cite{Hoaglin1985}). In considering such a class of elongation transformations to obtain a distribution, one is comparing the tail strength of the new distribution with that of the base distribution (such as a Gaussian or logistic). In this regard, one can think of tail strength or heavy-tailedness as an absolute concept, whereas the notion of elongation strength is a relative concept. In the following, we will first consider relative elongation compared to a base distribution for a generic random variable $W$. It should be clear that such a measure of relative tail behavior is independent of location and scale. Other properties, that such an elongation transform $T(\cdot)$, should satisfy are that it should preserve symmetry $T(w)\,{=}\,T(-w)$, and the base distribution should not be significantly transformed in the center, such that $T(w) \,{=}\, w \,{+}\, O(w^2)$ for $w$ around the mode. Then, to increase the tails of the resulting distribution relative to the base, it is important to assume that $T$ is strictly monotonically increasing transform that is convex, that is, one has the transform satisfying for positive $w>0$ that $T'(w) > 0$ and $T''(w) >0$. One such transformation family satisfying these properties includes the Tukey transformations.

\begin{definition}[Tukey transformations]\label{DefnTukey}
Consider a Gaussian random variable\break ${W\,{\sim}\,\mathrm{Normal}(0,1)}$ and transformation $X = r(w)$ then the resultant transformed loss random variable $X$ will be from a Tukey law if the corresponding transform $r(W)$ is given by
\begin{align}
X = r(W) = W\,T(W)^{\theta},
\end{align}
for a parameter $\theta \in \mathbb{R}$. Under this transformation we also have directly in closed form the quantile function of the loss random variable $X$ in terms of the quantile function of the based random variable $W$ as follows 
\begin{equation}
Q_{X}(\alpha; )  = a + b Q_{W}(\alpha)T\left(Q_{W}(\alpha)\right)^{\theta}
\end{equation}
with scaling and translation constants $a$, $b$.
\end{definition}

Typically, in several application settings, it will be desirable when working with such severity models to enforce a constraint that the tails of the resulting distribution after transformation are heavier than the Gaussian distribution. In this case, one should consider a transformation $T(w)$, which is positive, symmetric, and strictly monotonically increasing for positive values of $w \geq 0$. In addition, it will be desirable to obtain this property of heavy tails relative to the Gaussian to also consider setting the parameter $\theta \geq 0$. As discussed, a series of kurtosis transformations were proposed in the literature. The Tukey $h$, $k$, and $j$ transforms are provided in Definition~\ref{DefnTukeyhkj}.

\begin{definition}[Tukey's kurtosis transformations, $h$, $k$ and $j$ types] \label{DefnTukeyhkj}
The  $h$-type of transformation, denoted by $T_h(w)$, is given by
\begin{align}
T_h(w) = \exp\left(w^2\right).
\end{align}
The  $k$-type of transformation, denoted by $T_k(w)$, is given by
\begin{align}
T_k(w) = 1 + w^2.
\end{align}
The  $j$-type of transformation, denoted by $T_j(w)$, is given by
\begin{align}
T_j(w) = \frac{1}{2}\left[ \exp(w) + \exp(-w) \right].
\end{align}
\end{definition}

In addition to kurtosis transformations, there are skewness transforms that have been developed in the Tukey family, such as the g-transform.

\begin{definition}[Tukey's skewness transformation] \label{DefnTukeyg}
The  $g$-type of transformation, denoted by $T_g(w)$, is given by
\begin{align}
T_g(w) = \frac{\exp(w) - 1}{w}
\end{align}
The generalized $g$-type transform, denoted by $T^*_g(w)$, is given by
\begin{align}
T^*_g(w) = \left[1+c\frac{1-\exp\left(-g W\right)}{1+\exp\left(-g W\right)}\right]
\end{align}
\end{definition}
To nest all these transformations within one class of transformations, the work of \cite{fischer2010generalized} proposed a power series representation denoted by the subscript \textit{a} given in Equation (\ref{EqnPowerSeries}). This suggestion, though it nested the other families of distributions, is not practical for use as it involves the requirement of estimating a very large (infinite) number of parameters $a_i$ to obtain the data-generating mechanism:
\begin{align} \label{EqnPowerSeries}
T_a(w) = \sum_{i=0}^{\infty} a_i w^{2i}.
\end{align}
It was further observed in \cite{fischer2010generalized} that this nesting structure may be replaced with a different form, given by the general transformation taking the form given in Equation (\ref{EqnGeneralTran}):
\begin{align}\label{EqnGeneralTran}
T_{hjk}(w;\alpha,\beta,\gamma) = \left(1 + \frac{\left(w^2 + \gamma \right)^{\alpha} - \gamma^{\alpha} }{\beta} \right)^{\beta}, \;\;\;\; \alpha > 0, \beta \geq 1, \gamma > 0.
\end{align}
Then it is clear that the original $h$, $k$, and $j$ transformations are recovered with $T_h(w) = T_{hjk}(w;1,\infty,\gamma)$, $T_k(w) = T_{hjk}(w; 1,1,\gamma)$, and $T_j(w) \approx T_{hjk}(w; 0.5,\infty,0.5)$. Next, we explain the properties of specific subfamilies of distributions, showing how these results are derived for the basic g-and-h family and the g-and-k family.

\subsection{Examples of the g-and-h, g, h, and h--h Loss Models}
The g-and-h family can be considered as composed of three transformations that can produce subfamilies of non-Gaussian distributions for loss amount severity based on the g-distributions, the h-distributions, and the g-and-h distributional families. The basic specifications in which $g$ and $h$ components are treated as constants are given in Definition \ref{DefnGHfamily} in terms of transformations of a random variable $W$, typically Gaussian.

\begin{definition}[g-and-h Distributional family]\label{DefnGHfamily}
Let $W \,{\sim}\, \mathrm{Normal}(0,1)$ be a standard Gaussian random variable. Then the loss random variable $X$ has severity distribution given by the g-and-h distribution with parameters $a,b,g,h \in \mathbb{R}$, denoted $X \sim \mathrm{GH}(a,b,g,h)$, if $X$ is given by (for $g \neq 0$)
\begin{align} \label{EqnGANDH}
X = a + bWT_{g,h}\left(W; g,h\right)^{\theta} := a + bG(W)H(W)W,
\end{align}
where $\theta=1$ and
\begin{equation}
\label{eq:g_func}
G(w) = \frac{\exp(gw) - 1}{gw}
\end{equation}
and
\begin{equation}
\label{eq:h_func}
H(w) = \exp\left(\frac{hw^{2}}{2}\right).
\end{equation}
\end{definition}

One can observe that $a$ and $b$ account for location and scale, respectively. It can be checked from \eqref{eq:g_func} that the reshaping function $G$ is bounded from below by zero, that it is either monotonically increasing or monotonically decreasing for $g$ being, respectively, positive or negative, and that by rewriting it as its series expansion,
\begin{equation}
\label{eq:g_ex}
G(w) = 1 + \frac{gw}{2!} + \frac{(gw)^{2}}{3!} + \frac{(gw)^{3}}{4!} + \cdots,
\end{equation}
$G$ is equal to one at zero for all $g$. Thus $G$ generates asymmetry by scaling $w$ differently for different side of zero via the parameter $g$. Furthermore, as $G(w; g) = G(-w; -g)$, the sign of $g$ affects only the direction of skewness. For $g = 0$, by equation \eqref{eq:g_ex}, the constant function $G(w) = 1$ is obtained, and thus the symmetry remains unmodified. For $h > 0$, $H$ is a strictly convex even function with $H(0) = 1$, and thus it generates heavy tails by scaling upward the tails of $z$ while preserving the symmetry. When $h = 0$, the transformation given by Equation \eqref{EqnGANDH} generates the subfamily of g-distributions, which coincides with the family of shifted log-normal distributions for $g > 0$. When $g = 0$, the transformation generates the subfamily of h-distributions, which is symmetric and has heavier tails than normal distributions. The parameters $a$ and $b$ are linear transformations whereas the parameters $g$ and $h$ can be significantly extended to polynomials as discussed later, and play an important role in the skewness and kurtosis properties of the g-and-h family.

\begin{remark}
In general, one may consider the constants $g$ and $h$ to be more flexibly selected as polynomials, which would include higher orders of $W^2$. These polynomials could take the form, for example, of any integers $p$ and $q$ :
\begin{align}
\begin{split}
&g(w):=  \alpha_0 + \alpha_1 w + \cdots + \alpha_pw^{\,p},\\
&h(w):=  \beta_0 + \beta_1 w + \cdots + \beta_q w^q.
\end{split}
\end{align}
The addition of these polynomial terms can provide additional degrees of freedom to improve the ability to fit data. These have been shown to be significant when modeling certain types of loss data, as demonstrated by \cite{dutta2006tale} and \cite{peters2006bayesian}.
\end{remark}

Within this family of g-and-h distributions, one can also define the subfamilies of distributions given by the $g$ and the $h$ families. Again, we present these models in their simplest form, with constant $g$ or $h$, though in practice one may include polynomials in $W$ for such models.

\begin{definition}[g Distributional Family] \label{DefnGfamily}
Let $W \sim \mathrm{Normal}(0,1)$ be a standard Gaussian random variable. Then the loss random variable $X$ has severity distribution given by the $g$ distribution with parameters $a,b,g \in \mathbb{R}$, denoted $X \sim G(a,b,g)$, if $X$ is given by (for $g \neq 0$)
\begin{align}
X = a + b T_{g}\left(W;g\right) := a + b\frac{\exp\left(g W\right) - 1}{g}.
\end{align}
\end{definition}

\begin{remark}
Note that the g-distribution subfamily corresponds (in the case that $g$ is a constant) to a scaled LogNormal distribution.
\end{remark}

\begin{definition}[h Distributional Family]\label{DefnHfamily}
Let $W \,{\sim}\, \mathrm{Normal}(0,1)$ be a standard Gaussian random variable. Then the loss random variable $X$ has severity distribution given by the h distribution with parameters $a,b,h \in \mathbb{R}$, denoted $X \sim H(a,b,h)$, if $X$ is given by
\begin{align}
X = a + bWT_{h}\left(W; a,b,h\right) := a + bW\exp\left( \frac{h W^2}{2}\right).
\end{align}
\end{definition}

In addition, one may obtain an asymmetric class of $h$--$h$ distributions studied by \cite[section~2.2]{morgenthaler2000fitting}, \cite{headrick2012characterizing} and \cite{headrick2012logistic}. The asymmetric $h$--$h$ distribution transformation is given in Definition \ref{DefnHHdist}.

\begin{definition}[Double $h$--$h$ Distributional Family] \label{DefnHHdist}
Let $W \sim \mathrm{Normal}(0,1)$ be a standard Gaussian random variable. Then the loss random variable $X$ has severity distribution given by the unit h--h distribution with parameters $h_l,h_r \in \mathbb{R}$, denoted $X \sim HH(h_l,h_r)$, if $X$ is given~by
\begin{align}
X = a + b W T_{h,h}\left(W; h_l,h_r\right) :=
\begin{cases}
a + b W \exp\left( \dfrac{1}{2} h_l W^2 \right), & W \leq 0, \\[11pt]
a + b W \exp\left( \dfrac{1}{2} h_r W^2 \right), & W \geq 0, \\
\end{cases}
\end{align}
for $h_r \geq 0$ and $h_l \geq 0$.
\end{definition}

To conclude this section, there is also a generalized g-and-h family that is given in Definition \ref{DefnGenGHfamily}, see discussions in \cite{RaynerMacGillivray2002}.

\begin{definition}[Generalized g-and-h Distributional family]\label{DefnGenGHfamily}
Let $W \,{\sim}\, \mathrm{Normal}(0,1)$ be a standard Gaussian random variable. Then the loss random variable $X$ has severity distribution given by the g-and-h distribution with parameters $a,b,g,h \in \mathbb{R}$, denoted $X \sim Generalized-\mathrm{GH}(a,b,g,h)$, if $X$ is given by (for $g \neq 0$)
\begin{align} \label{EqnGANDHgen}
X = a + bWT^*_{g,h}\left(W; g,h\right)^{\theta} := a + bG^*(W)H(W)W,
\end{align}
where $\theta = 1$ and
\begin{equation}
\label{eq:geng_func}
G^*(w) = \left[1+c\frac{1-\exp\left(-g W\right)}{1+\exp\left(-g W\right)}\right]
\end{equation}
and
\begin{equation}
\label{eq:genh_func}
H(w) = \exp\left(\frac{hw^{2}}{2}\right).
\end{equation}
\end{definition}

\subsection{Examples of the g-and-k and g-and-j Loss Models}

The g-and-k family of loss models, as parameterized in \cite{haynes2005bayesian} is given by combining the $g$ and the $k$ transforms as given in Definition \ref{DefnGandK}.

\begin{definition}[g-and-k Distributional family]\label{DefnGandK}
Let $W \,{\sim}\, \mathrm{Normal}(0,1)$ be a standard Gaussian random variable. Then the loss random variable $X$ has severity distribution given by the g-and-k distribution with parameters $a,b,g,k \in \mathbb{R}$, denoted $X \sim GK(a,b,g,k)$, if $X$ is given by (for $g \neq 0$)
\begin{align} \label{EqnGANDK}
X = a + b W T_{g,k}\left(W; a,b,g,k\right)^{\theta} &:= a + bG^*(W)K(W)W
\end{align}
where $\theta = 1$ and 
\begin{equation}
G^*(w) = \left[1+c\frac{1-\exp\left(-g W\right)}{1+\exp\left(-g W\right)}\right]
\end{equation}
and
\begin{equation}
K(w) = \left(1+W^2\right)^k.
\end{equation}
with $a \in \mathbb{R}$ is location, $b > 0$ is scale, $g \in \mathbb{R}$ is the skewness measure, $k > -0.5$ is a measure of kurtosis and $c$ is a constant.
\end{definition}

Similarly, the g-and-j family of loss models is obtained by combining the $g$ and the $j$ transforms as given in Definition \ref{DefnGandJ}.

\begin{definition}[g-and-j Distributional family]\label{DefnGandJ}
Let $W \,{\sim}\, \mathrm{Normal}(0,1)$ be a standard Gaussian random variable. Then the loss random variable $X$ has severity distribution given by the g-and-k distribution with parameters $a,b,g,j \in \mathbb{R}$, denoted $X \sim GJ(a,b,g,k)$, if $X$ is given by (for $g \neq 0$)
\begin{align} \label{EqnGANDJ}
X = a + b W T_{g,j}\left(W; a,b,g,j\right)^{\theta} &:= a + bG^*(W)J(W)W
\end{align}
where $\theta = 1$ and 
\begin{equation}
G^*(w) = \left[1+c\frac{1-\exp\left(-g W\right)}{1+\exp\left(-g W\right)}\right]
\end{equation}
and
\begin{equation}
J(w) = \frac{1}{2}\left[\exp(W)+\exp(-W)\right].
\end{equation}
with $a \in \mathbb{R}$ is location, $b > 0$ is scale, $g \in \mathbb{R}$ is the skewness measure, and in this case one can set $j = 1$.
\end{definition}

In the following sections we will explore properties of the two more widely used families of models the g-and-h, and its sub-families, as well as the g-and-k claims severity models.

\section{g-and-h, g, h, h--h, and g-and-k Family Distribution and Density Functions} 

In this section we discuss properties of the Tukey sub-families of loss models and in particular different ways that people have sought to evaluate and present the distribution and density functions for the popular sub-families such as the g-and-h, generalized g-and-h and g-and-k families. In general, it will be informative for this section to remind the reader of the following basic property.

\begin{proposition} \label{PropTukeyResults}
If $X$ is a continuous random variable distributed according to distribution ${X\,{\sim}\,F(x)}$, which is monotonically increasing on support $\text{Supp}\left\{F(x)\right\} = \left\{ x: 0 < F(x) < 1\right\}$, then, in this general case, one can show that the quantile function $Q_X(\alpha) = F^{-1}(\alpha)$ for $\alpha \in [0,1]$ determines the relationship between the random variable $X$ and any other continuous random variable with monotonically increasing distribution, say $W \sim G(w)$, through the relationship given as follows
\begin{align} \label{EqnRel}
Q_X(\alpha) = F^{-1}\left(G\left(Q_W(\alpha)\right)\right).
\end{align} 
Furthermore, the following relationship between the quantile function of a random variable $X$ and its density cab be obtained by using the identity for differentiation of an inverse function given by
\begin{equation}
\frac{d}{dx}g^{-1}(x) = \left[\frac{1}{dx}g\left(g^{-1}(x)\right)\right]^{-1}
\end{equation}
this result when applied to the quantile function of random variable $X$ produces the following relationship
\begin{equation} \label{EqnDensityIdent}
\frac{dQ_X(\alpha)}{d\alpha} = \frac{1}{f_X\left(Q_X(\alpha)\right)}
\end{equation}
where $Q_X(\alpha)$ is the quantile function for random variable $X$ at quantile level $\alpha$ and $f_X(\cdot)$ represents the density for random variable $X$. One can then also apply this to the relationship in Equation \ref{EqnRel} to obtain
\begin{equation} 
\begin{split}
\frac{dQ_X(\alpha)}{d\alpha} &= \frac{d}{du}Q_X(u)\frac{du}{d\alpha} \\
&=\frac{g\left(Q_W(\alpha)\right)}{f_X\left(F^{-1}\left(G\left(Q_W(\alpha)\right)\right)\right)}.
\end{split}
\end{equation}
\end{proposition}

In the remainder of this paper we will consider to utilise the most popular choice of reference distribution in the literature which refers to the standard Gaussian base distribution, i.e. \\ $W\sim G(w) = \Phi(w;0,1)$. We note that it is however trivial to modify the results below for other choices of distribution. In this Gaussian case, one can show that for any continously differentiable transformation $X = T(W)$, $X$ will have a density given in Equation (\ref{EqnDensityGHtran}) with respect to the standard Gaussian density $\phi(\cdot)$. 
\begin{align}\label{EqnDensityGHtran}
f_X(x) = \frac{\phi\left(T^{-1}(x)\right)}{ T'\left( T^{-1}(x) \right) }.
\end{align}
In this case, one can also observe that when the transform $T(\cdot)$ increases rapidly, the resulting density is heavy-tailed. For instance, conversely a slower linear growth in the function $T(\cdot)$ results in tail behavior for the distribution of random variable $X$ being equivalent to a Gaussian

These two general results in Proposition \ref{PropTukeyResults} can then been used to characterized the distribution and density functions for different members of the Tukey family of quantile specified loss models. In the following results we basically apply the same methodology to obtain the density and distribution for each of the different Tukey classes of loss model. We begin with the density of the superclass of transformations presented previously according to the quantile transformation $T_{hjk}$. 

\begin{definition}[Super Class $T_{hjk}$ Density]
One can state the following basic properties for the loss random variable $X=r(W)=WT_{hjk}(W)^{\theta}$, the loss density  $f_X(\cdot)$ and quantile functions $Q_X(\cdot)$, for loss random variable $X$, are given by
\begin{align}
f_X(x; h,j,k) &= \frac{1}{Q'_{X}\left( Q^{-1}_X\left(x\right) \right) }  \notag\\
&= \frac{ \phi\left(r^{-1}(x)\right) }{ r'\left( r^{-1}(x) \right) },\;\;\;
\inf\left\{x: x\in S\right\} < x < \sup\left\{x: x\in S\right\}\\
Q_X(\alpha) &= r\left(Q_W(\alpha)\right), \;\; \alpha \in [0,1],\notag
\end{align}
with $S$ the appropriate support of the random variable $X$ and
\begin{align*}
r'(w) = T_{hjk}(w)^{\theta - 1}\left( T_{hjk}(w) + \theta w T'_{hjk}(w) \right).
\end{align*}
\end{definition}
Clearly, this density representation is a composition of two functions, one of which can only be evaluated typically numerically due to general non-closed from expressions for the inversion.

In an analogous manner one can ofcourse then find the distribution and density for the other Tukey families of loss model. The first observation one can make for the g-and-h family is that since the transformations are monotonically increasing as long as $h > 0$, the quantile function of the g-and-h distribution is readily available. Let $x_{u}$ denoted the level $u \in [0,1]$ quantile function of a g-and-h distribution,
\begin{equation}
\label{eq:gh_qtlfun}
x(u; a, b, g, h) =
\begin{cases}
a + b \frac{\exp(g z_{u}) - 1}{g} \exp\left(\frac{h z_{u}^{2}}{2}\right) & \text{if $g \neq 0$} \\
a + b z_{u} \exp\left(\frac{h z_{u}^{2}}{2}\right) & \text{if $g = 0$}
\end{cases}
\end{equation}
where $z_{u}$ is the $u$-level quantile of the standard normal distribution, $a \in \mathbb{R}$, $b \in (0, \infty)$, $g \in \mathbb{R}$, and $h \in [0, \infty)$ are parameters responsible for location, scale, asymmetry, and heavy-tailedness, respectively. This result was utilised in \cite{degen2007quantitative} and \cite{headrick2008parametric} to obtain expressions for the density as such a composite function.

\begin{definition}[g-and-h Distribution Function (constant $g$ and $h$ with $h > 0$)]\label{DefnGandHDist} Consider the g-and-h distributed random variable $X \sim \mathrm{GH}(a=0,b=1,g,h)$ with constant parameters $g$ and $h > 0$. The distribution function can be specified according to the following composite function:
\begin{align}
F_X(x; g,h) &= \Phi\left(r^{-1}(x)\right),\\
f_X(x; g,h) &= \frac{1}{Q'_{X}\left( Q^{-1}_X\left(x\right) \right) },  \notag\\
&= \frac{ \phi\left(r^{-1}(x)\right) }{ r'\left( r^{-1}(x) \right) },
\end{align}
where $\Phi(\cdot)$ is the standard Gaussian distribution and the function $r(x)$ is specified by
\begin{align}
r(x) = \frac{ \exp\left(g x\right) -1 }{ g} \exp\left( \frac{h x^2}{2}\right).
\end{align}
where the derivative is then given by
\begin{align}
\begin{split}
r'(x) &= \frac{d}{dx}\left[\frac{\exp\left(g x\right) - 1}{g} \exp\left( \frac{h x^2}{2}\right)\right] \\
&= \exp\left( gx + \frac{hx^2}{2} \right) + \frac{h}{g}x \exp\left( \frac{h x^2}{2} \right)\left( \exp(g x) - 1\right).
\end{split}
\end{align}
\end{definition}

In this parametrization, the parameter $g$ will control the skew of the distribution both in terms of the sign and the magnitude, while the parameter $h$ will control heaviness of the tails and is related directly to the kurtosis. This will be discussed further when the regular variation properties of this model are explored. As demonstrated previously, the original Tukey h-type transformation had $\theta = 1$ and an addition scaling of $\frac{1}{2}$. This transformation has the property that its derivative
\begin{align}
\frac{d}{dx}r(x) = \left( 1 + hx^2\right)\exp\left(\frac{1}{2}h x^2 \right) \geq 1
\end{align}
for all $h \geq 0$. In addition, in the following discussions, it will be useful to recall the following properties of the g-and-h family of distributions:
\begin{enumerate}
\item{The g-and-h transformation can be shown to be strictly monotonically increasing in its argument, that is, for all $w_1 \leq w_2$ one has $T_{gh}(w_1) \leq T_{gh}(w_2)$;}
\item{If $a {\,=\,} 0$, then the g-and-h transformation satisfies the condition $T_{-g,h}(W) {\,=\,} -T_{g,h}(-W)$.}
\end{enumerate}

In the case of the generalized g-and-h distribution one has the Tukey quantile transform producing loss random variable $X$ according to based random variable $W$ given by
\begin{align}
T^*_{gh}(W) = \left[1+c\frac{1-\exp\left(-g W\right)}{1+\exp\left(-g W\right)}\right]\exp\left( \frac{h W^2}{2}\right),
\end{align}
with $r(W) = WT(W)^{\theta}$ and typically in this family we also consider $\theta = 1$. This then produces the following density and distributions.

\begin{definition}[Generalized g-and-h Distribution and Density Functions]\label{DefnGandHDistGen} Consider the generalized g-and-h distributed random variable $X \sim \mathrm{GH}(a=0,b=1,g,h)$ with constant parameters $g$ and $h > 0$. The distribution function can be specified according to the following composite function:
\begin{align}
F_X(x; g,h) &= \Phi\left(r^{-1}(x)\right),\\
f_X(x; g,h) &= \frac{1}{Q'_{X}\left( Q^{-1}_X\left(x\right) \right) },  \notag\\
&= \frac{ \phi\left(r^{-1}(x)\right) }{ r'\left( r^{-1}(x) \right) },
\end{align}
where $\Phi(\cdot)$ is the standard Gaussian distribution and the function $r(x)$ is specified by
\begin{align}
r(x) = \left[1+c\frac{1-\exp\left(-g x\right)}{1+\exp\left(-g x\right)}\right]x\exp\left( \frac{h x^2}{2}\right),
\end{align}
where the derivative is then given by
\begin{align}
\begin{split}
 r'(x) &:= \frac{d}{dx}\left\{\left[1+c\frac{1-\exp\left(-g x\right)}{1+\exp\left(-g x\right)}\right]x\exp\left( \frac{h x^2}{2}\right)\right\} \\
&=\frac{\exp(h x^2/2)\left(1 + \exp(g x)\right)^2(1 + hx^2) + 2c\exp(g x)\left(gx +(1 + hx^2)\sinh(gx)\right)}{\left(1 + \exp(gx)\right)^2}.
\end{split}
\end{align}
\end{definition}

In the case of the g-and-k distribution family one has the quantile function given by
\begin{align}
b(W) = WT_{gk}(W)^{\theta} = \left[1+c\frac{1-\exp\left(-g W\right)}{1+\exp\left(-g W\right)}\right] W\left(1+W^2\right)^k.
\end{align}

\begin{definition}[g-and-k Distribution and Density Functions] \label{DefnGKDensDist}
Consider the generalized g-and-k distributed random variable $X \sim \mathrm{GH}(a=0,b=1,g,k)$ with constant parameters $g$ and $k > -0.5$. The distribution function can be specified according to the following composite function:
\begin{align}
F_X(x; g,k) &= \Phi\left(k^{-1}(x)\right),\\
f_X(x; g,k) &= \frac{1}{Q'_{X}\left( Q^{-1}_X\left(x\right) \right) },  \notag\\
&= \frac{ \phi\left(r^{-1}(x)\right) }{ r'\left( r^{-1}(x) \right) },
\end{align}
where $\Phi(\cdot)$ is the standard Gaussian distribution and the function $r(x)$ is specified by
\begin{align}
r(x) = \left[1+c\frac{1-\exp\left(-g w\right)}{1+\exp\left(-g w\right)}\right] w\left(1+w^2\right)^k,
\end{align}
where the derivative is then given by
\begin{align}
\begin{split}
r'(w;g,k) &:= \frac{d}{dw}\left\{\left[1+c\frac{1-\exp\left(-g w\right)}{1+\exp\left(-g w\right)}\right] w\left(1+w^2\right)^k\right\}\\
&=\frac{2\exp(gw)\exp(1+w^2)^k\left(1+cgw + 2kw^2 + (1 + 2kw^2)\text{Cosh}(gw) + c(1+2kw^2)\text{Sinh}(gw)\right)}{\left(1+\exp(gw)\right)^2}
.
\end{split}
\end{align}
\end{definition}

\section{g-and-h, g, h, h--h, and g-and-k Family Statistical Properties of Relevance to Claims Modelling} 

One advantage of the specifications presented above, of the distribution and density functions with regard to a particular quantile function, is that the statistical properties of these distributions can now be easily studied. For instance, the mode and moments of the distribution can be characterized. The result in Proposition \ref{PropModeGH} provides the mode for the g-and-h, generalized g-and-h and the g-and-k distributions.

\begin{proposition}[Mode of the g-and-h, generalized g-and-h and g-and-k Densities] \label{PropModeGH}
Consider the g-and-h distributed random variable $X \sim \mathrm{GH}(a=0,b=1,g,h)$ with constant parameters $g$ and $h > 0$, the generalized g-and-h given by $X \sim \text{Generalized}-\mathrm{GH}(a=0,b=1,g,h,c)$ and the g-and-k distribution $X \sim GK(a=0,b=1,g,k)$ with constant parameters $g$ and $k > -0.5$. In each of these models, which will be generically denoted here by transform $X = r(W)$, the mode of the density is located at the value $\widetilde{w} = \text{Mode}\left[W\right]$, which produces a maximum value of the densities at $ f_{r(W)}(\widetilde{w})$, depending on the transform $r(w)$ in each case, and can be found as the solution to the following equations when $w = \widetilde{w}$, which is selected to satisfy
\begin{align}
\frac{d}{dw}\left[ \frac{f_W(w)}{r'(w)}\right] = 0,
\end{align}
\end{proposition}

Analogously the medians can also be obtained.

\begin{proposition}[Median of the g-and-h, generalized g-and-h and g-and-k Densities] \label{PropMedGH}
Consider the g-and-h distributed random variable $X \sim \mathrm{GH}(a=0,b=1,g,h)$ with constant parameters $g$ and $h > 0$, the generalized g-and-h given by $X \sim \text{Generalized}-\mathrm{GH}(a=0,b=1,g,h,c)$ and the g-and-k distribution $X \sim GK(a=0,b=1,g,k)$ with constant parameters $g$ and $k > -0.5$. In each of these models, which will be generically denoted here by transform $X = r(W)$, the median of the density is located at the value $w_{0.5} = \text{Median}\left[W\right]$ and will correspond to the median being the limit of\; $\lim_{w \rightarrow 0}r(w)=0$.  
\end{proposition}

\begin{remark}
Therefore, one sees that in each of the g-and-h, generalized g-and-h and g-and-k distributions the median of the data set will be the parameter $a$. Furthermore, in the case of the h-type and double h-type Tukey distributions, the median and mode are at the origin (for $a=0$).
\end{remark}

One can also obtain the moments of Tukey family of distributions, with generically denoted Tukey quantile transform given by $r(W) = WT(W)^{\theta}$, as the solution to the following integrals, where the $n$-th moment is given with respect to the transformed moments of the base density as follows:
\begin{equation}
\mathbb{E}\left[X^n\right] = \mathbb{E}\left[r(W)^n\right] = \int\limits_{-\infty}^{\infty} r(w)^n f_W(w) dw,
\end{equation}

From such a result, one may now express the moments of the g-and-h, generalized g-and-h and g-and-k distributed random variables according to the results in Proposition \ref{PropMomentsGH}, Proposition \ref{PropMomentsGenGH} and Proposition \ref{PropMomentsGK}. 

Before presenting these we note the following from \cite{dutta2002measuring} that since the g-distribution is a horizontally shifted LogNormal distribution, then the moments of the g-distribution take the same form as those of a LogNormal model with appropriate adjustment for the translation. The h-distributional family is symmetric (except the double h--h family); consequently, all odd-order moments  for the h-subfamily are zero.

\begin{proposition}[Moments of the g-and-h Density] \label{PropMomentsGH}
Consider the g-and-h distributed random variable $X \sim \mathrm{GH}(a=0,b=1,g,h)$ with constant parameters $g$ and $h > 0$. The $n$-th integer moment is given with respect to the standard Normal distribution and the $n$-th power of the transformed quantile function given by
\begin{align}
r(W) = a + b\frac{ \exp\left(g W\right) -1 }{ g} \exp\left( \frac{h W^2}{2}\right).
\end{align}
to produce moments according to the relationship
\begin{align}
\mathbb{E}\left[X^n\right] = \mathbb{E}\left[r(W)^n\right]
\end{align}
which will exist if $h \in \left[0,\frac{1}{n}\right)$.  One can also observe more generally that under the g-and-h transform the following identity holds with regard to powers of the standard Gaussian, $W \sim \mathrm{Normal}(0,1)$, such that
\begin{align}
\begin{split}
X^n &= r(W)^n = T_{g,h}(W; a,b,g,h)^n \\
&= \left(a + bT_{g,h}(W; a=0,b=1,g,h) \right)^n \\
&= \sum_{i=0}^n \frac{n!}{(n-i)! i!} a^{n-i}b^i T_{g,h}(W; a=0,b=1,g,h)^i,
\end{split}
\end{align}
which will produce moments given by
\begin{align}
\begin{split}
\mathbb{E}\left[X^n\right] &= \mathbb{E}\left[\left(a + bT_{g,h}(W; a=0,b=1,g,h) \right)^n\right]\\
&= \sum_{i=0}^n \frac{n!}{(n-i)! i!} a^{n-i}b^i \mathbb{E}\left[T_{g,h}(W; a=0,b=1,g,h)^i\right].
\end{split}
\end{align}
Furthermore, it was shown by \cite{dutta2002measuring} that when it exists one can obtain the general expression
\begin{align}
\mathbb{E}\left[T_{g,h}(W; a=0,b=1,g,h)^i\right] = \frac{ \sum_{r=0}^i (-1)^r \frac{i!}{(i-r)! r!}\exp\left( \frac{(i-r)^2g^2}{2(1-ih)} \right)  }{ \sqrt{(1-ih)}g^i },
\end{align}
\end{proposition}

The results in Proposition \ref{PropMomentsGH} then produce the following four population moments for the basic g-and-h loss model in closed form for $a = 0$ and $b=1$:
\begin{align*}
%
\mathbb{E}\left[X\right] &  = \left[ \exp\left( \frac{g^2}{2 - 2h} \right)
- 1\right]\left[ g\sqrt{1-h}\right]^{-1}\\
\mathbb{E}\left[X^2\right] & = \left[ 1 - 2\exp\left( \frac{g^2}{2-4h} \right)
+ \exp\left( \frac{2g^2}{1 - 2h} \right)\right]\left[ g^2\sqrt{1-2h}\right]^{-1}\\
\mathbb{E}\left[X^3\right]& 
= \left[ 3\exp\left( \frac{g^2}{2 - 6h} \right) + \exp\left( \frac{9g^2}{2 - 6h} \right)-\, 3\exp\left( \frac{2g^2}{1 - 3h} \right) - 1\right]\left[ g^3\sqrt{1-3h}\right]^{-1}\\
\mathbb{E}\left[X^4\right]& = s(g,h)\exp
\left( \frac{8g^2}{1 - 4h} \right) \left[ g^4\sqrt{1-4h}\right]^{-1}.
\end{align*}
with the function $s(g,h)$ being given by
\begin{align*}
 s(g,h) \,{=}\, \left(\!1\,{+}\,6 \exp\left(\!\frac{6g^2}{4h - 1}\!\right)\,{+}\,\exp\left(\!\frac{8g^2}{4h-1}\!\right)
 \,{-}\,4 \exp\left(\!\frac{7g^2}{8h-2}\!\right)\,{-}\,4\exp\left(\!\frac{15g^2}{8h-2}\!\right)\!\right).
\end{align*}

Analagously then we can also find the $n$-th order integer moments for the generalized g-and-h and the g-and-k models as follows.

\begin{proposition}[Moments of the Generalized g-and-h Density] \label{PropMomentsGenGH}
Consider the Generalized g-and-h distributed random variable $X \sim \mathrm{GH}(a=0,b=1,g,h)$ with constant parameters $g$ and $h > 0$. The $n$-th integer moment is given with respect to the standard Normal distribution and the $n$-th power of the transformed quantile function
\begin{align}
r(W) = a + b\left[1+c\frac{1-\exp\left(-g W\right)}{1+\exp\left(-g W\right)}\right]W\exp\left( \frac{h W^2}{2}\right).
\end{align}
given by
\begin{align}
\mathbb{E}\left[X^n\right] = \mathbb{E}\left[r(W)^n\right]
\end{align}
which will exist if $h \in \left[0,\frac{1}{r}\right)$. Hence, one obtains moments given by
\begin{align}
\begin{split}
\mathbb{E}\left[X^n\right] &= \mathbb{E}\left[\left(a + bT^*_{g,h}(W; a=0,b=1,g,h) \right)^n\right]\\
&= \sum_{i=0}^n \frac{n!}{(n-i)! i!} a^{n-i}b^i \mathbb{E}\left[T^*_{g,h}(W; a=0,b=1,g,h)^i\right].
\end{split}
\end{align}
The moments of the generalized transform can not in general be obtained in closed form except in some special cases. However, one make the following McClaren series expansion of the term $\left[G^*(W)H(W)/W\right]^i$ and then approximate the moments as follows at say pth order. We provide the result for 3rd order series expansion of the transform below
{\small{
\begin{align} 
\begin{split}
&\left(T^*_{g,h}(W; a=0,b=1,g,h)\right)^i \\
&= W^i\left(1 + \frac{1}{2}icgW + \left(\frac{ih}{2} + \frac{1}{8}c^2g^2(i-1)i\right)W^2 +
\left(\frac{1}{4}i^2cgh - \frac{1}{24}icg^3 + \frac{1}{48}c^3g^3(i-2)(i-1)i\right)x^3 + O(x^4)\right).
\end{split}
\end{align}
}}
This can then be integrated to produce the approximate $i$-th order moments given by
\begin{equation}
\begin{split}
&3\pi 2^{5 - \frac{i}{2}}\mathbb{E}\left[T^*_{g,h}(W; a=0,b=1,g,h)^i\right] \\
& = cgi \left[12 (2 + h i (2 + i)) + g^2 (2 + i) (-2 + c^2 (2 - 3 i + i^2))\right] \Gamma\left(1 + \frac{i}{2}\right)\\ 
&	\;\;\;\; + 3 \sqrt{2} \left[c^2 g^2 i (-1 + i^2) + 4 (2 + h i (1 + i))\right] \Gamma\left(\frac{1 + i}{2}\right) \\
&	\;\;\;\; + (-1)^i \left[-c g i (12 (2 + h i (2 + i)) + g^2 (2 + i) (-2 + c^2 (2 - 3 i + i^2)))\right] \Gamma\left(1 + \frac{i}{2}\right) \\
& \;\;\;\; + (-1)^i \left[ 3 \sqrt{2} (c^2 g^2 i (-1 + i^2) + 4 (2 + h i (1 + i)))\right]\Gamma\left(\frac{1 + i}{2}\right)  + O(x^{4+i+1})
\end{split}
\end{equation}
\end{proposition}

Similarly to the results obtained above we can obtain the moments of the g-and-k as follows.

\begin{proposition}[Moments of the g-and-k Density] \label{PropMomentsGK}
Consider the g-and-k distributed random variable $X \sim GK(a=0,b=1,g,k)$ with constant parameters $g$ and $k > -0.5$. The $n$-th integer moment of the distribution is given with respect to the standard Normal distribution and the $n$-th power of the transformed quantile function
\begin{align}
r(W) = \left[1+c\frac{1-\exp\left(-g W\right)}{1+\exp\left(-g W\right)}\right] W\left(1+W^2\right)^k,
\end{align}
given by
\begin{align}
\mathbb{E}\left[X^n\right] = \mathbb{E}\left[r(W)^n\right].
\end{align}
Hence, one obtains moments given by
\begin{align}
\begin{split}
\mathbb{E}\left[X^n\right] &= \mathbb{E}\left[\left(a + bT^*_{g,k}(W; a=0,b=1,g,k) \right)^n\right]\\
&= \sum_{i=0}^n \frac{n!}{(n-i)! i!} a^{n-i}b^i \mathbb{E}\left[T^*_{g,k}(W; a=0,b=1,g,k)^i\right].
\end{split}
\end{align}
The moments of the G-and-K transform can not in general be obtained in closed form except in some special cases. However, one may make the following McClaren series expansion and then approximate the moments as follows at say p-th order. We provide the result for 3rd order series below
\begin{align} 
\begin{split}
&\left(T^*_{g,k}(W; a=0,b=1,g,k)\right)^i \\
&= W^i (g W)^i \left[\frac{1}{2^{i+1}\pi} - \frac{1}{3\pi}\left(\frac{1}{2^{3+i}}(6 + g^2i -12k)\right)W^2  + O(W^4)\right].
\end{split}
\end{align}
This can then be integrated to produce the approximate $i$-th order moments given by
\begin{equation}
\mathbb{E}\left[T^*_{g,h}(W; a=0,b=1,g,h)^i\right] \approx -\frac{\left[\Gamma\left(1/2 + i\right) \left((-1)^i (-g)^i + g^i\right) \left((g^2 - 12 k) i (1 + 2 i) - 12 \right) \right]}{24\sqrt{2}\pi}.
\end{equation}
\end{proposition}

\begin{remark}
These results allow one to perform model estimation via moment matching of model moments to empirical moments of the loss data.
\end{remark}

Furthermore, using these moment identities one can easily then find the skew, kurtosis, and coefficient of variations for the g-and-h, generalized g-and-h and g-and-k loss distribution models as well as the subfamilies for the g-distributions and h-distributions. 

In addition, there are numerous authors who have studied the generalized properties of quantile-based functionals of asymmetry and kurtosis (see examples in Definition \ref{DefnGenSkewKurt}; also see \cite{balanda1990kurtosis}, \cite{rayner2002numerical}, and \cite{balanda1988kurtosis}).

\begin{definition}[Generalized Skewness and Kurtosis Functionals] \label{DefnGenSkewKurt}
In considering the generalizations of the skewness and kurtosis for transformation-based quantile function severity models, one can utilize the generalized specifications given for the skewness functional, for a given distribution $F_X(x)$ with respect to its quantile function $Q_X(x)$ by
\begin{align}
\gamma_F = \frac{ Q_X(\alpha) + Q_X(1-\alpha) - 2Q_X\left( \frac{1}{2}\right) }{ Q_X(\alpha) - Q_X(1-\alpha)},  \quad \alpha \in (0,1).
\end{align}
In addition, there is the spread functional given by
\begin{align}
S_F = Q_X(\alpha) - Q_X(1-\alpha), \quad \alpha \in (0,1).
\end{align}
\end{definition}

Such measures were discussed by \cite{balanda1990kurtosis} and it can be shown that $\left| \gamma_F(\alpha) \right| \leq 1$. In the case of the g-and-h family of severity models, one would obtain the forms given in Definition \ref{DefnGenSkewSpread}.

\begin{definition}[Generalized Skewness and Kurtosis for g-and-h and generalized g-and-h Families] \label{DefnGenSkewSpread}
Consider the g-and-h distributed random varaible $X \sim \mathrm{GH}(a=0,b=1,g,h)$ with constant parameters $g$ and $h > 0$ then the generalized skewness and kurtosis are given for the g-and-h model according to expressions
\begin{align*}
S_F &= Q_X(\alpha) - Q_X(1-\alpha) \\
&= \frac{ \exp\left(g \Phi^{-1}(\alpha)\right) -1 }{ g} \exp\left( \frac{1}{2}h \Phi^{-1}(\alpha)^2\right) \\
&\quad - \frac{ \exp\left(g \Phi^{-1}(1-\alpha)\right) -1 }{ g} \exp\left( \frac{1}{2}h \Phi^{-1}(1-\alpha)^2\right),\\
\gamma_F &= \frac{ Q_X(\alpha) + Q_X(1-\alpha) - 2Q_X\left( \frac{1}{2}\right) }{ Q_X(\alpha) - Q_X(1-\alpha) }  \\
&=\frac{ \frac{ \exp\left(g \Phi^{-1}(\alpha)\right) -1 }{ g} \exp\left( \frac{1}{2}h \Phi^{-1}(\alpha)^2\right) }{S_F} + \frac{\frac{ \exp\left(g \Phi^{-1}(1-\alpha)\right) -1 }{ g} \exp\left( \frac{1}{2}h \Phi^{-1}(1-\alpha)^2\right) }{S_F} \\
&\quad - 2 \frac{\frac{ \exp\left(g \Phi^{-1}(0.5)\right) -1 }{ g} \exp\left( \frac{1}{2}h\Phi^{-1}
(0.5)^2\right)}{S_F}. %
\end{align*}
In the case of the generalized g-and-h model they are given according to expressions
{\small{
\begin{align*}
S^*_F &= Q_X(\alpha) - Q_X(1-\alpha) \\
&= \left[1+c\frac{1-\exp\left(-g \Phi^{-1}(\alpha)\right)}{1+\exp\left(-g \Phi^{-1}(\alpha)\right)}\right]\Phi^{-1}(\alpha)\exp\left( \frac{h \Phi^{-1}(\alpha)^2}{2}\right)\\
&\quad - \left[1+c\frac{1-\exp\left(-g \Phi^{-1}(1-\alpha)\right)}{1+\exp\left(-g \Phi^{-1}(1-\alpha)\right)}\right]\Phi^{-1}(1-\alpha)\exp\left( \frac{h \Phi^{-1}(1-\alpha)^2}{2}\right) ,\\
\gamma^*_F &= \frac{ Q_X(\alpha) + Q_X(1-\alpha) - 2Q_X\left( \frac{1}{2}\right) }{ Q_X(\alpha) - Q_X(1-\alpha) }  \\
&=\frac{\left[1+c\frac{1-\exp\left(-g \Phi^{-1}(\alpha)\right)}{1+\exp\left(-g \Phi^{-1}(\alpha)\right)}\right]\Phi^{-1}(\alpha)\exp\left( \frac{h \Phi^{-1}(\alpha)^2}{2}\right) }{S_F} 
+ \frac{\left[1+c\frac{1-\exp\left(-g \Phi^{-1}(1-\alpha)\right)}{1+\exp\left(-g \Phi^{-1}(1-\alpha)\right)}\right]\Phi^{-1}(1-\alpha)\exp\left( \frac{h \Phi^{-1}(1-\alpha)^2}{2}\right) }{S_F} \\
&\quad - 2 \frac{\left[1+c\frac{1-\exp\left(-g \Phi^{-1}(1/2)\right)}{1+\exp\left(-g \Phi^{-1}(1/2)\right)}\right]\Phi^{-1}(1/2)\exp\left( \frac{h \Phi^{-1}(1/2)^2}{2}\right)}{S_F}. %
\end{align*}
}}
\end{definition}

Analogously one can trivially find the generalized skewness and kurtosis for the g-and-k and g-and-j families.

\subsection{Tail Properties of the g-and-h and g-and-k Loss Models}
In terms of the tail behavior of the g-and-h family of distributions, the properties of such severity models have been studied by numerous authors such as \cite{morgenthaler2000fitting} and \cite{degen2007quantitative}. In particular, the tail property (index of regular variation) for the g-and-h family of distributions was first studied for the h-distribution by \cite{morgenthaler2000fitting} and later for the g-and-h distribution by \cite{degen2007quantitative} (see Proposition \ref{PropGHRV}). In addition, the second-order regular variation properties of the g-and-h family of distributions was studied by \cite{degen2007quantitative}.

In order to study the properties of regular variation of the g-and-h family of loss distribution models it is first important to recall some basic definitions. First, we note that a postive measurable function $f(\cdot)$ is regularly varying if it satisfies the conditions in Definition \ref{DefnRVFn}, see discussion in \cite{karatzas1991brownian}.

\begin{definition}[Regularly Varying Function]\label{DefnRVFn}
A positive measurable function $f(\cdot)$ is regularly varying (at infinity) with an index $\alpha \in \mathbb{R}$ if it satisfies:
\begin{itemize}
\item{It is defined on some neighbourhood $[x_0,\infty)$ of infinity; and
}
\item{It satisfies the following limiting relationship
\begin{align}
\lim_{x \rightarrow \infty} \frac{f(\lambda x)}{f(x)} = \lambda^{\alpha}, \;\; \forall \lambda > 0.
\end{align}
}
\end{itemize}
\end{definition}
We note that when $\alpha = 0$, then the function $f(\cdot)$ is said to be slowly varying (at infinity). From this definition one can show that a random variable has a regularly varying distribution if it satisfies the condition in Definition \ref{RegVarDefn}.

\begin{definition}[Regularly Varying Random Variable] \label{RegVarDefn}
A loss random variable $X$ with distribution $F_X(x)$ taking positive support is said to be regularly varying with index $\alpha \geq 0$ if the right tail distribution $\overline{F}_X(x) = 1 - F_X(x)$ is regularly varying with index $-\alpha$.
\end{definition}

The following important features can be noted about regularly varying distributions as shown in Theorem \ref{ThmRVDistsProp}, see detailed discussion in \cite{bingham1989regular}.

\begin{theorem}[Properties of Regularly Varying Distributions]\label{ThmRVDistsProp}
Given a loss distribution $F_X(x)$ satisfying $F_X(x) < 1$ for all $x \geq 0$, the following conditions on $F_X(x)$ can be used to verify that it is regularly varying such that $F_X(x) \in RV_{\alpha}$:
\begin{itemize}
\item{If $F_X(x)$ is absolutely continuous with density $f_X(x)$ such that for some $\alpha > 0$ one has the\break limit\vspace*{3pt}
\begin{align}
\lim_{x \rightarrow \infty} \frac{x f_X(x)}{\overline{F}_X(x)} = \alpha.
\end{align}
Then $f_X(x)$ is regularly varying with index $-(1 + \alpha)$ and consequently $\overline{F}_X(x)$ is regularly varying with index $-\alpha$;
}
\item{If the density $f_X(x)$ for loss distribution $F_X(x)$ is assumed to be regularly varying with index $-(1 +\alpha)$ for some $\alpha > 0$. Then the following limit,\vspace*{3pt}
\begin{align}
\lim_{x \rightarrow \infty} \frac{x f_X(x)}{\overline{F}_X(x)} = \alpha,
\end{align}
will also be satisfied if $\overline{F}_X(x)$ is regularly varying with index $-\alpha$ for some $\alpha > 0$ and the density $f_X(x)$ will be ultimately monotone.}
\end{itemize}
\end{theorem}

Many additional properties are described for such heavy tailed distribution and density functions. Here we will utilise the above stated conditions to assess the regular variation properties of the right tail of the g-and-h family of loss models. In particular we will see if a single distributional parameter characterizes the heavy tailed feature as captured by the notion of regular variation index, or if the relationship is more complex.

\begin{proposition}[Index of Regular Variation of g-and-h Distribution] \label{PropGHRV}
Consider the random variable $W \sim \mathrm{Normal}(0,1)$ and a loss random variable $X$, which has severity distribution given by the g-and-h distribution with parameters $a,b,g,h \in \mathbb{R}$, denoted $X \sim \mathrm{GH}(a,b,g,h)$, with $h > 0$ and density (distribution) $f(x)$ $($and $F(x))$ . Then the index of regular variation is obtained by considering the following limit
\begin{align}
\lim_{x \rightarrow \infty} \frac{x f(x)}{\overline{F}(x)} = \lim_{x\rightarrow\infty} \frac{ \phi(u)\left(\exp(g u) - 1\right) }{ \left( 1 - \Phi(u) \right)\left( g\exp(gu) + hu(\exp(gu) -1) \right) } = \frac{1}{h}\\ \nonumber
\end{align}
for $u = k^{-1}(x)$ where the function $k(x)$ is given by
\begin{align}
k(x) = \frac{ \exp\left(g x\right) -1 }{ g} \exp\left( \frac{h x^2}{2}\right).\\ \nonumber
\end{align}
Hence, one can state that $\overline{F} \in RV_{-\frac{1}{h}}$.
\end{proposition}

The asymptotic tail behavior of the h-family of Tukey distributions was studied by \cite{morgenthaler2000fitting} and is given in Proposition \ref{PropHTypeTail}.

\begin{proposition}[h-Type Tail Behaviour] \label{PropHTypeTail}
Consider the h-type transformation, where $W \sim \mathrm{Normal}(0,1)$ is a standard Gaussian random variable and the loss random variable $X$ has severity distribution given by the h-distribution with parameters $a,b,h \in \mathbb{R}$, denoted $X \sim H(a,b,h)$ according to
\begin{align}
X = T_{h}\left(W; a,b,h\right) := a + bW\exp\left( \frac{h W^2}{2}\right).\\ \nonumber
\end{align}
Then the asymptotic tail index of the h-type distribution is then given by ${1}/{h}$. This is equivalent to the g-and-h family for $g \neq 0$.
\end{proposition}

This shows that the h-type family has a Pareto heavy-tailed property, hence the restriction that moments will only exist on the order of less than ${1}/{h}$. The g-family of distributions can be shown to be subexponential in the tail behavior but not regularly varying. It was shown  \cite[theorem~2.2]{degen2007quantitative} that one can obtain an explicit form for the function of slow variation in the g-and-h family as detailed in Theorem \ref{ThmGHSV}.

\begin{theorem}[Slow Variation Representation of g-and-h Severity Models] \label{ThmGHSV}
Consider the\break random variable $W \sim \mathrm{Normal}(0,1)$ and a loss random variable $X$, which has severity distribution given by the g-and-h distribution with parameters $a,b,g,h \in \mathbb{R}$, denoted $X \sim \mathrm{GH}(a,b,g,h)$, with $g > 0$ and $h > 0$ and density (distribution) $f(x)$ $($and $F(x))$ . Then $\overline{F}(x) = x^{-{1}/{h}} L(x)$ for some slowly varying function $L(x)$ given as $x \rightarrow \infty$ by
\begin{align}
L(x) = \frac{h}{ \sqrt{2 \pi} g^{{1}/{h}} } \frac{ \left[\exp\left( \frac{g}{h}\sqrt{ g^2 + 2h\ln(gx) } - \frac{g^2}{h} \right) - 1\right]^{{1}/{h}}  }{\sqrt{ g^2 + 2h\ln(gx) } - g} \left(1 + O\left( \frac{1}{\ln x} \right) \right).
\end{align}
\end{theorem}

From this explicit Karamata representation developed by \cite{degen2007quantitative}, it was also shown that one can obtain the second-order regular variation properites of the g-and-h family.

The implications of these findings are that the g-and-h distribution, under the parameter restrictions $g>0$ and $h >0$, belongs to the domain of attraction of an Extreme Value Distribution, such that $X \sim \mathrm{GH}(a,b,g,h)$ with distribution $F$ satisfying $F \in MDA\left(H_{\gamma}\right)$ where $\gamma = h > 0$. As a consequence, by the Pickands--Balkema--de Haan Theorem, discussed in detail in companion book \cite{cruz2015fundamental}, one can state that there exists an Extreme Value Index (EVI) constant $\gamma$ and a positive measurable function $\beta(\cdot)$ such that the following result between the excess distribution of the g-and-h (denoted by $F_u(x) = \mathbb{P}\mathrm{r}\left( X-u \leq x | X > u\right)$ and the generalized Pareto distribution (GPD) is satisfied in the tails
\begin{align}
\lim_{u \uparrow \infty} \sup_{x \in (0, \infty)} \left| F_u(x) - G_{\gamma,\beta(u)}(x) \right| = 0.
\end{align}
For discussion on the rate of convergence in the tails, see \cite{raoult2003rate} and the application of this theorem to the g-and-h case by \cite{degen2007quantitative} where it is shown that the order of covergence is given by $O\left( A\exp\left(V^{-1}(u)\right) \right)$ for functions
\begin{align}
\begin{split}
V(x) &:= \overline{F}^{-1}\left( \exp(-x)\right),  \\
A(x) &:= \frac{V''(\ln x) }{V'(\ln x)} - \gamma.
\end{split}
\end{align}
Hence, the conclusion from this analysis regarding the tail convergence of the excess distribution of the g-and-h family toward the GPD $G_{\gamma,\beta(u)}(x)$ is given explicitly by
\begin{align}
\frac{ \ln L(x)}{\ln x} \sim \sqrt{2}\frac{ g }{ h^{\frac{3}{2} } } \frac{1}{\sqrt{\ln(x)}} = O\left( \frac{1}{\sqrt{\ln\left(  k^{-1}(x)\right)}}\right), \;\; x \rightarrow \infty.
\end{align}

\begin{remark}
The implications of this slow rate of convergence are that when data for severities are obatained from a loss process, if a goodness-of-fit test suggests that one may not reject the null hypothesis that these data came from a g-and-h distribution, then one should avoid performing estimation of the extreme quantiles, such as those used to measure the capital via the Value-at-Risk, via methods based on Peaks Over Threshold (POT) or Extreme Value Theory (EVT) based penultimate approximations.
\end{remark}

\begin{proposition}[Index of Regular Variation of the generalized g-and-h Distribution] 
Consider the random variable $W \sim \mathrm{Normal}(0,1)$ and a loss random variable $X$, which has severity distribution given by the g-and-h distribution with parameters $a,b,g,h \in \mathbb{R}$, denoted $X \sim \mathrm{GH}(a,b,g,h)$, with $k > 0$ and density (distribution) $f(x)$ $($and $F(x))$ . 
Recall that we had for the generalized g-and-h loss model the function $r(x)$ with $a=0$ and $b=1$ given by
\begin{align}
r(x) = \left[1+c\frac{1-\exp\left(-g x\right)}{1+\exp\left(-g x\right)}\right]x\exp\left( \frac{h x^2}{2}\right).
\end{align}
Using this, we can then find the index of regular variation at $x \rightarrow \infty$ as obtained by considering the following limit
\begin{align}
\lim_{x \rightarrow \infty} \frac{x f(x)}{\overline{F}(x)} &= \lim_{x\rightarrow\infty} \frac{x \phi\left(r^{-1}(x)\right)}{r'\left(r^{-1}(x)\right)\left[1 - \Phi\left(r^{-1}(x)\right)\right]} \\
\end{align}
which by defining $u = r^{-1}(x)$ and further noting that $u(1-\Phi(u))/\phi(u) \rightarrow 1$ as $x \rightarrow \infty$ one can rewrite as follows using the reciprocal law of limits that states that 
if $\lim_{x \rightarrow c}g(x) = M$ and $M \neq 0$ then $\lim_{x \rightarrow c}\frac{1}{g(x)} = \frac{1}{M}$ as well as the product law of limits that
\begin{align}
\begin{split}
&\lim_{u \rightarrow \infty} \frac{\phi(u)r(u)}{(1-\Phi(u))r'(u)} \\
&= \lim_{u \rightarrow \infty} \frac{\phi(u)r(u)}{(1-\Phi(u))r'(u)}\\
&= \lim_{u \rightarrow \infty} \frac{u r(u)}{r'(u)} \\
&= \lim_{u \rightarrow \infty} \frac{\left(1 + \exp(gu)\right)u^2}{1 + gu + hu^2 + \exp(gu)(1 + hu^2)}
\end{split}
\end{align}
Now in the case that $g < 0$ then one has that $exp(gu) \rightarrow 0$ as $u \rightarrow \infty$ and hence one obtains
\begin{align}
\begin{split}
&\lim_{u \rightarrow \infty} \frac{\phi(u)r(u)}{(1-\Phi(u))r'(u)} \\
&= \lim_{u \rightarrow \infty} \frac{\left(1 + \exp(gu)\right)u^2}{1 + gu + hu^2 + \exp(gu)(1 + hu^2)} = 1/h
\end{split}
\end{align}
Where, as expected we again see that in the geralized g-and-h model it produces a distribution that satisfies $\overline{F} \in RV_{-\frac{1}{h}}$.
\end{proposition}

This is not unexpected since the $G$ transform in each case drives the skewness and not the kurtosis. We can also obtain this analysis for the g-and-k model, this yields that the g-and-k does not admit a finite limit in either sign of the parameter $g$, showing that such a model is not regularly varying, as we see in the case of the g-and-h models. However, even though this is the case we can still assess the relative heavy tailedness of the g-and-k models compared to the base distribution under the Tukey k-transform.

\section{Estimating the General Tukey Family Loss Model Parameters}
\label{sec:estimation}
Several studies in the statistics literature have been performed on the estimation of these types of quantile function specified models, see likelihood based estimation in \cite{RaynerMacGillivray2002} and \cite{hossain2009numerical} or the Bayesian approaches such as in \cite{haynes2005bayesian},\cite{peters2006bayesian} and \cite{allingham2009bayesian}. In this section we propose and develop a class of novel estimation methods based on L-moments for a range of Tukey families which shows favorable properties compared to previously proposed methods. Before presenting this new approach developed in this paper we first comment on a few approaches previously proposed for instance in the g-and-h family.

\subsection{Estimating the g-and-h Loss Model Parameters}
As the g-and-h family does not admit a closed-form density function, the likelihood function can only be expressed in terms of the inverse quantile function;
\begin{equation}
\label{eq:gh_like}
f_{\mathrm{gh}}(x_{1} \ddd x_{n};\boldsymbol{\theta}_{\mathrm{gh}}) = \prod_{i=1}^{n}\frac{d}{dx_{i}}X^{-1}(x_{i};\boldsymbol{\theta}_{\mathrm{gh}}),
\end{equation}
where $x_{1} \ddd x_{n}$ are observations, $\boldsymbol{\theta}_{\mathrm{gh}} = (a, b, g, h)'$ is the parameter vector, and $X^{-1}$ is the inverse quantile function. The high computational cost of evaluating the likelihood function comes from the fact that $X^{-1}$ can not be expressed in closed-form, and thus the quantile function must be inverted using an iterative root-search algorithm. The maximum likelihood (ML) estimates of the g-and-h parameters can be found by iteratively searching over the parameter space. The quality of the ML estimates is investigated via simulations by \cite{RaynerMacGillivray2002} for quantile distribution families generated using skewness and spread functionals, where the authors find that the ML method can be unstable for small samples.

Another method of fitting the g-and-h distributions is by matching moments. The $k$-th moment exists if and only if $h \in [0, \frac{1}{k})$. The expression of the $k$-th raw moment can be found in \citeA{MartinezIglewicz1984}; for $a = 0$, $b = 1$, and $g \ne 0$, the $k$-th raw moment is given by
\begin{equation}
\label{eq:gh_rmom}
m^{*}_{k} = \E\left(x^{k}\right) = \frac{1}{g^{k}\sqrt{1-kh}} \sum_{i=0}^{k}(-1)^{i}\binom{k}{i}\exp\left(\frac{[(k-i)g]^{2}}{2(1-kh)}\right),
\end{equation} 
and for $g = 0$,
\begin{equation}
\label{eq:h_rmom}
m^{*}_{k} = \E\left(x^{k}\right) =
\begin{cases}
\frac{k!}{2^{k/2}[(k/2)!]}(1-kh)^{-(n+1)/2} & \text{if $k$ is even} \\
0 & \text{if $k$ is odd}.
\end{cases}
\end{equation}
Given the $k$-th raw moment, the $k$-th central moment can be computed by,
\begin{equation}
\label{eq:gh_cmom}
m_{k} = \E\left[\left(x-m^{*}_{1}\right)^{k}\right] = \sum_{i=0}^{k}\binom{k}{i}(-1)^{k-i}m^{*}_{i}m_{1}^{k-i}.
\end{equation}
From the central moments, the skewness $\zeta_{3}$ and kurtosis $\zeta_{4}$ are given by
\begin{equation}
\label{eq:gh_skewkurt}
\begin{aligned}
\zeta_{3} &= m_{3} / m_{2}^{3/2}, \\
\zeta_{4} &= m_{4} / m_{2}^{2}.
\end{aligned}
\end{equation}
As the skewness and kurtosis are location and scale invariant shape measures, $g$ and $h$ can be simultaneously found by minimising the objective,
\begin{equation}
\label{eq:mom_gnh}
(\zeta_{3} - \hat{\zeta}_{3})^{2} + (\zeta_{4} - \hat{\zeta}_{4})^{2},
\end{equation}
subject to $0 \le h < \frac{1}{4}$, where $\hat{\zeta}_{3}$ is the sample skewness and $\hat{\zeta}_{4}$ is the sample kurtosis. Given the estimates of $g$ and $h$, $b$ and $a$ can be solved straightforwardly as follows.
\begin{equation}
\label{eq:mom_bna}
\begin{aligned}
b &= \sqrt{\hat{m}_{2} / m_{2}}, \\
a &= \hat{m}_{1} - bm_{1},
\end{aligned}
\end{equation}
where $\hat{m}_{1}$ is the sample mean and $\hat{m}_{2}$ is the sample variance. The moment matching estimator is proposed by \citeA{HeadrickKowalchukSheng2008}, however the quality of such estimator is not investigated in depth by the authors.

Instead of matching moments, the g-and-h parameters can be estimated by matching quantiles, as proposed by \cite{XuIglewiczChervoneva2014}. Let $0 < u_{1} < \cdots < u_{q} < 1$ denote a set of quantile levels chosen a priori, $X(u_{1}) \ddd X(u_{q})$ denote the set of g-and-h quantiles, and $\hat{\chi}_{u_{1}} \ddd \hat{\chi}_{u_{q}}$ denote the set of sample quantiles. The estimate of $\boldsymbol{\theta}_{\mathrm{gh}}$ is found by minimising the objective,
\begin{equation}
\label{eq:qm}
\sum_{i=1}^{q}[X(u_{i}) - \hat{\chi}_{u_{i}}]^{2},
\end{equation}
subject to $b > 0$ and $h \ge 0$. The quality of the quantile matching estimator is determined by the selection of $u_{1} \ddd u_{q}$. \cite{XuIglewiczChervoneva2014} choose equally spaced quantiles given by $u_{i} = \frac{i-1/3}{q+1/3}$, for $i \in \{1 \ddd q\}$. By treating $u_{1} \ddd u_{q}$ as auxiliary parameters, the number of quantiles $q \in \{4 \ddd 20\}$ is then selected by minimising the Akaike information criterion (AIC) given by
\begin{equation}
\label{eq:aic}
\mathrm{AIC} = n \log\left(\frac{\mathrm{SSE}}{n}\right) + 2(q + 1),
\end{equation}
where
\begin{equation}
\label{eq:sse}
\mathrm{SSE} = \sum_{i=1}^{n}[X(p_{i}) - \hat{\chi}_{p_{i}}]^{2}
\end{equation}
and
\begin{equation}
\label{eq:p}
p_{i} = \frac{i - 1 / 3}{n + 1 / 3}.
\end{equation}
Notice that, for each $q$, the corresponding $\mathrm{SSE}$ is computed using the same number of quantiles as the number of observations, thus making use of the full sample.

\subsubsection{New Robust Estimation Approach for g-and-h Loss Models Based on Method of L-moments}
\label{subsec:mlmom}
We propose a method for fitting Tukey transform distributions such as the g-and-h family using L-moments, this is a general extension of previous specific modified Tukey family models of \cite{headrick2012logistic} as there is no assumption here on the choice of base line distribution for $W$.
Some of the advantages that L-moments have over conventional moments are that they exist whenever the mean of the distribution exists and in addition they are nearly unbiased for all sample sizes and distributions, and they have been noted to be more robust in the presence of outliers, see discussions in \cite{hosking2005regional} and \cite{hodis2012power}.

L-moments are defined by \citeA{Hosking1990} to be certain linear combinations of expectations of order statistics. Specifically, let $x_{(1)} \le x_{(2)} \le \cdots \le x_{(n)}$ denote a sample of ordered observations. For $k \in \{1, 2, \,\ldots\}$, the $k$-th L-moment is defined as
\begin{equation}
\label{eq:lmom_sum}
l_{k} = \frac{1}{k}\sum_{i=0}^{k-1}(-1)^{i}\binom{k-1}{i}\E\left[x_{(k-i)}\right].
\end{equation}
The connection between L-moments and a quantile function becomes apparent when L-moments are expressed as projections of a quantile function onto a sequence of orthogonal polynomials that forms a basis of $L^{2}$;
\begin{equation}
\label{eq:lmom_int}
l_{k} = \int_{0}^{1} F^{-1}_X(u)L_{k-1}(u) \, du,
\end{equation}
where $L_{k}$ is the $k$-th shifted Legendre polynomial in the sequence given generically by
\begin{equation}
L_{k-1}(u) = \sum_{j=0}^{k-1}\frac{(-1)^{k-j}(k+j)!}{(j!)^2(k-j)!} u^j
\end{equation}
Compared to classical moments, L-moments are able to characterise a wider range of distributions as all L-moments of a distribution exist if and only if the mean exits. Furthermore, a distribution with finite mean is uniquely characterised by its sequence L-moments. Using the representation of \eqref{eq:lmom_int}, the first four L-moments are given by
\begin{equation}
\label{eq:lmom}
\begin{aligned}
l_{1} &= \int_{0}^{1} F^{-1}_X(u) \, du, \\
l_{2} &= \int_{0}^{1} F^{-1}_X(u)(2u - 1) \, du, \\
l_{3} &= \int_{0}^{1} F^{-1}_X(u)(6u^{2} - 6u + 1) \, du, \\
l_{4} &= \int_{0}^{1} F^{-1}_X(u)(20u^{3} - 30u^{2} + 12u - 1) \, du.
\end{aligned}
\end{equation}

\begin{remark}
In the above notation, the $F^{-1}_X(u)$ is understood to be the quantile function of the random variable $X$ at quantile level $u$. Hence, for the Tukey family of models the L-moments are to be considered with respect to the integral of the transform of the quantile function of the base distribution, which is implicitly included above when we write $F^{-1}_X(u)$ and could be considered with respect to the base distribution quantile function as follows $F^{-1}_X(u) :=r\left(F^{-1}_W(u)\right)$.
\end{remark}
 
The location and scale invariant L-moment ratios, $\tau_{3}$ and $\tau_{4}$, analogous to the classical skewness and kurtosis, respectively termed L-skewness and L-kurtosis in \citeA{Hosking1990}, are defined as
\begin{equation}
\begin{aligned}
\tau_{3} &= l_{3} / l_{2}, \\
\tau_{4} &= l_{4} / l_{2}.
\end{aligned}
\end{equation}
Unlike the classical skewness and kurtosis, L-skewness and L-kurtosis are bounded, with $\tau_{3} \in (-1, 1)$ and $\tau_{4} \in [\frac{1}{4}(5 \tau_{3}^{2} - 1), 1)$ for continous based distributions. The boundedness of L-moment ratios makes them easy to interpret.

The sample L-moments, also known as L-statistics, are unbiased estimates of L-moments based on the order statistics of an observed sample. In particular, the first four sample L-moments are given by
\begin{equation}
\label{eq:lstat}
\begin{aligned}
\hat{l}_{1} &= \hat{M}_{0}, \\
\hat{l}_{2} &= 2\hat{M}_{1} - \hat{M}_{0}, \\
\hat{l}_{3} &= 6\hat{M}_{2} - 6\hat{M}_{1} + \hat{M}_{0}, \\
\hat{l}_{4} &= 20\hat{M}_{3} - 30\hat{M}_{2} + 12\hat{M}_{1} - \hat{M}_{0},
\end{aligned}
\end{equation}
where $\hat{M}_{k}$ is the $k$-th sample probability weighted moment \cite{GreenwoodEtAl1979}, given by
\begin{equation}
\label{eq:pwm}
\hat{M}_{k} =
\begin{cases}
\frac{1}{n}\sum_{i=1}^{n}x_{(i)} & \text{if $k = 0$} \\
\frac{1}{n}\sum_{i=1}^{n}\frac{(i - 1)(i - 2) \cdots (i - k)}{(n - 1)(n - 2) \cdots (n - k)} x_{(i)} & \text{if $k > 0$}.
\end{cases}
\end{equation}
An alternative, but numerically equivalent, method of computing the sample L-moments is by following closely the definition in \eqref{eq:lmom_sum}. See \citeA{Wang1996} for details.

The most general approach to performing Method of L-Moments that will be applicable for any base distribution $W$ and any Tukey sub-family of models involves matching L-moments of the population to the sample L-moments. We note that in general, the integrals in \eqref{eq:lmom} for the Tukey families of models may be obtained accurately via an one-dimensional numerical integration algorithm such as the adaptive quadrature. However, in some cases we can also obtain closed form expressions for these L-Moments as detailed below where we derive these moments in closed form for the Gaussian based distribution most commonly used in practice. To achieve the L-moment expressions we will work with the quantile of the Gaussian base function for $W$ given for the standard normal by
\begin{equation}
F^{-1}_W(u) =  \sqrt{2}\text{erf}^{-1}\left(2u-1\right).
\end{equation}

In the case of the first L-moment we can find a closed form expression for the Tukey families of models and in the case of higher order L-moments we will utilise the series expansion to find a result for the L-moments to any desired accuracy by truncation of the series expansion and explicit integrations as follows.	We will first consider to make a change of variable given by
\begin{equation}
R = \sqrt{2}\text{erf}^{-1}\left(2u-1\right),
\end{equation}
which means that we have the $n$-th L-moment for the general Tukey family of models which can be written according to the expression
\begin{equation}
\begin{split}
l_n &= \sum_{j=0}^{n-1} \frac{(-1)^{n-j}(n+j)!}{(j!)^2(n-j)!} \int_{0}^{1} r\left(F^{-1}_W(u)\right) u^j \, du,\\
&= \sum_{j=0}^{n-1} \frac{\sqrt{2}(-1)^{n-j}(n+j)!}{(j!)^2(n-j)!} \int_{-\infty}^{\infty} r\left(R\right)\left(\frac{1}{2}(\Phi(R) + 1)\right)^j \phi\left(\sqrt{2}R\right)dR 
\end{split}
\end{equation}
where we used the fact that
\begin{equation}
\frac{d}{du}\left[\text{erf}^{-1}\left(u\right)\right] = \frac{1}{2}\sqrt{\pi}\exp\left[\left(\text{erf}^{-1}\left(u\right)\right)^2\right],
\end{equation}
and $\text{erf}^{-1}\left(-1\right) = -\infty$ and $\text{erf}^{-1}\left(1\right) = \infty$. Furthermore, for the first for L-Moments we can identify the Legendre polynomials for this change of variable as follows
\begin{equation} \label{eq:lmom2}
\begin{aligned}
L_{0}(R) &= 1, \\
L_{1}(R) &= \Phi(R), \\
L_{2}(R) &= 3\Phi(R)^2 + 3\Phi(R) + 1,\\
L_{4}(R) &= 10\Phi(R)^3 + 15\Phi(R)^2 + 12\Phi(R) - 6,
\end{aligned}
\end{equation}
which means we can write the first four L-Moments as follows
\begin{equation}
\begin{aligned}
l_{1} &=  \sqrt{2} \int_{-\infty}^{\infty} r\left(R\right) \phi\left(\sqrt{2}R\right)dR,\\
l_{2} &=  \sqrt{2} \int_{-\infty}^{\infty} r\left(R\right)\Phi(R) \phi\left(\sqrt{2}R\right)dR,\\
l_{3} &=  \sqrt{2} \int_{-\infty}^{\infty} r\left(R\right) \left[3\Phi(R)^2 + 3\Phi(R) + 1\right] \phi\left(\sqrt{2}R\right)dR, \\
l_{4} &=  \sqrt{2} \int_{-\infty}^{\infty} r\left(R\right) \left[10\Phi(R)^3 + 15\Phi(R)^2 + 12\Phi(R) - 6\right] \phi\left(\sqrt{2}R\right)dR.
\end{aligned}
\end{equation}
These representations under the change of variable will now be used in the following propositions to obtain the L-Moments for each of the different Tukey sub-families considered.
	
\begin{proposition}[g-Family Loss Model Population L-Moments]\label{PropG} {
The first four population L-Moments of the g-family of Tukey transform Loss models is given for $a=0$, $b=1$ by considering $$r(R) = \frac{\exp(gR) - 1}{g},$$ giving:
\begin{equation} \label{eq:lmomg}
\begin{aligned}
l_{1} &=\frac{\exp\left(\frac{g^2}{2}\right)-1}{g},\\
l_{2} &=  \frac{\sqrt{2\pi}}{2g}\exp\left(g^2\right) - \frac{1}{2g} +\frac{2}{\sqrt{2\pi}}\psi_1\left(1, \frac{1}{2}, \frac{3}{2}, \frac{1}{2}; - \frac{1}{2}, \frac{g^2}{4} \right),\\
l_{3} &=  \frac{3}{g\sqrt{\pi}} \psi_1\left(1, \frac{1}{2}, \frac{3}{2}, \frac{1}{2}; - \frac{1}{2}, \frac{g^2}{4} \right) + \frac{3\sqrt{2\pi}}{4g} \exp\left(\frac{g^2}{4}\right) - \frac{3\text{Arctan}\left(\sqrt{3}\right)}{g\pi\sqrt{\pi}} + 3l_2 + l_1,\\
l_{4} &= 5 l_3 + 27 l_2 - 1 l_1 + \frac{45\text{Arctan}\left(\sqrt{3}\right)}{g\pi\sqrt{\pi}} + \frac{30\sqrt{2}}{\sqrt{3}\pi}\text{Arctan}\left(\frac{1}{\sqrt{15}}\right) + O\left(\int_{0}^{\infty} R^{3}\Phi(R)^3 \phi\left(\sqrt{2}R\right)dR \right).
\end{aligned}
\end{equation}
}
\end{proposition}
The proof of this result can be obtained in Appendix A. Similarly, one may obtain the first four l-moments for the h-family, the g-and-h family and the g-and-k-family of Tukey elongation loss models are given respectively as follows.

\begin{proposition}[h-Family Loss Model Population L-Skewness and L-Kurtosis]\label{PropH}
{
The first four population L-Moments of the h-family of Tukey transform Loss models is given for $a=0$, $b=1$ by considering $$r(R) = R\exp\left(\frac{hR^2}{2}\right),$$ giving:
\begin{equation}
\begin{aligned}
l_{1} &=  \sqrt{2} \int_{-\infty}^{\infty} r\left(R\right) \phi\left(\sqrt{2}R\right)dR = \frac{1}{\sqrt{\pi}} \int_{-\infty}^{\infty} R \exp\left(-(1-h)R^2\right)dR = 0, \;\; h < 1,\\
l_{2} &=  \sqrt{2} \int_{-\infty}^{\infty} r\left(R\right)\Phi(R) \phi\left(\sqrt{2}R\right)dR= \frac{1}{\sqrt{\pi}} \frac{1}{(1-h)}\left(1 + \frac{1}{\sqrt{1+2(1-h)}}\right),\\
l_{3} &=  \sqrt{2} \int_{-\infty}^{\infty} r\left(R\right) \left[3\Phi(R)^2 + 3\Phi(R) + 1\right] \phi\left(\sqrt{2}R\right)dR = 3 l_2 + l_1, \\
l_{4} &=  \sqrt{2} \int_{-\infty}^{\infty} r\left(R\right) \left[10\Phi(R)^3 + 15\Phi(R)^2 + 12\Phi(R) - 6\right] \phi\left(\sqrt{2}R\right)dR\\
&= \frac{10}{8\sqrt{2\pi}} l_3 + 12 l_2 - 6 l_1 + 
\frac{30}{8\pi(1-h)\sqrt{\pi[1 + 2(1-h)]}} 
\text{Arctan}\left(\frac{1}{\sqrt{[2 + 4(1-h)][\frac{3}{2}+(1-h)]}}\right).
\end{aligned}
\end{equation}
}
\end{proposition}
The proof of this result can be obtained in Appendix A.

\begin{proposition}[k-family loss model population L-skewness and L-kurtosis]\label{PropK}{
The first four population L-Moments of the h-family of Tukey transform Loss models is given for $a=0$, $b=1$ by considering $$r(R) = R\left(1+R\right)^k,$$ giving:
\begin{equation}
\begin{aligned}
l_{1} &=  \sqrt{2} \int_{-\infty}^{\infty} r\left(R\right) \phi\left(\sqrt{2}R\right)dR = 0,\\
l_{2} &=  \sqrt{2} \int_{-\infty}^{\infty} r\left(R\right)\Phi(R) \phi\left(\sqrt{2}R\right)dR\\
&= \frac{1}{\sqrt{2\pi}} \left\{I_1 + \widetilde{I}_1 +2k I_3 + (k-1) k I_5 +1/3 (k-2) (k-1) k I_7 \right\} +O\left(\int_{-\infty}^{\infty}R^9 \Phi(R) \exp\left(-R^2\right)dR\right),\\
l_{3} &=  \sqrt{2} \int_{-\infty}^{\infty} r\left(R\right) \left[3\Phi(R)^2 + 3\Phi(R) + 1\right] \phi\left(\sqrt{2}R\right)dR \\
&= \frac{3}{2\sqrt{2\pi}} \left\{(G_1 + \widetilde{G}_1) + k (G_{3} + \widetilde{G}_3) + 1/2 (k-1) k (G_{5} + \widetilde{G}_5) + 1/6 (k-2) (k-1) k (G_{7}+\widetilde{G}_7) \right\} \\ 
&+ 3 l_2 + \left(\frac{3}{4\sqrt{\pi}} + 1\right)l_1 +O\left(\int_{-\infty}^{\infty}R^9 \Phi(R) \exp\left(-R^2\right)dR\right),\\
l_{4} &=  \sqrt{2} \int_{-\infty}^{\infty} r\left(R\right) \left[10\Phi(R)^3 + 15\Phi(R)^2 + 12\Phi(R) - 6\right] \phi\left(\sqrt{2}R\right)dR\\
&= \frac{5}{\sqrt{2}}\left[l_3 - 3l_2 - l_1 \right]  + 12 l_2 - 6 l_1 
+ \frac{30}{8\pi\sqrt{3\pi}} \text{Arctan}\left(\frac{1}{4}\right) +
O\left(\int_{-\infty}^{\infty}R^9 \Phi(R) \exp\left(-R^2\right)dR\right),
\end{aligned}
\end{equation}
with 
\begin{equation}
\begin{split}
&I_1 = \frac{1}{2}\left(1 + \frac{1}{\sqrt{3}}\right),\\
&I_{2m+1} = \frac{(-1)^m m!}{2} - \frac{(-1)^m}{2}\left.\frac{\partial^m}{\partial p^m}\left(\frac{1}{p\sqrt{p+1/2}}\right)\right|_{p=1},\\
&\widetilde{I}_1 = \frac{1}{2}\left(1 - \frac{1}{\sqrt{3}}\right),\\
&G_{2n+1} = \frac{(-1)^n}{\sqrt{2}\pi}\left. \frac{\partial^n}{\partial p^n}\left(\frac{1}{p\sqrt{1/2 + p}}\text{Arctan}\left(\frac{1}{\sqrt{1 + 2p}}\right) \right)\right|_{p=1},\\
&\widetilde{G}_{2n+1} = \frac{(-1)^{n+1}}{\sqrt{2}\pi}\left. \frac{\partial^n}{\partial p^n}\left(\frac{1}{p\sqrt{1/2 + p}}\text{Arctan}\left(\frac{-1}{\sqrt{1 + 2p}}\right) \right)\right|_{p=1}.
\end{split}
\end{equation}
}
\end{proposition}

Finally, one can also find the g-and-h family Population L-moments in closed form as follows.

\begin{proposition}[g-and-h-Family Loss Model Population L-Skewness and L-Kurtosis]\label{PropGH}{
The first four population L-Moments of the g-and-h family of Tukey transform Loss models is given for $a=0$, $b=1$ by considering $$r(R) = \frac{\exp(gR)-1}{g}\exp\left(\frac{hR^2}{2}\right), $$  giving:
\begin{equation}
\begin{aligned}
l_{1} &=  \sqrt{2} \int_{-\infty}^{\infty} r\left(R\right) \phi\left(\sqrt{2}R\right)dR\\
&= \frac{\exp\left(\frac{g^2}{4 - 2 h}\right)\sqrt{2}}{g\sqrt{2 - h}} - \frac{2 }{g\sqrt{4 - 2 h}},\;\;\;\; h < 2,\\
\end{aligned}
\end{equation}

\begin{equation}
\begin{aligned}
l_{2} &=  \sqrt{2} \int_{-\infty}^{\infty} r\left(R\right)\Phi(R) \phi\left(\sqrt{2}R\right)dR\\
&=  \frac{1}{2g\sqrt{\pi}} \frac{\exp\left(\frac{g^2}{(4 - 2 h)}\right) \sqrt{2 \pi}}{\sqrt{2 - h}} + \frac{1}{g\pi\sqrt{2}(1-h/2)}\psi_1\left(1, \frac{1}{2}, \frac{3}{2}, \frac{1}{2}; - \frac{1}{2(1-h/2)}, \frac{g^2}{4(1-h/2)} \right) - \frac{1}{g\sqrt{\pi}} \frac{\sqrt{2\pi}}{2},\\
l_{3} &=  \sqrt{2} \int_{-\infty}^{\infty} r\left(R\right) \left[3\Phi(R)^2 + 3\Phi(R) + 1\right] \phi\left(\sqrt{2}R\right)dR \\
&= 3l_2 + l_1 + \sum_{n=0}^{\infty} \frac{6(g)^{n-1}}{\sqrt{\pi}n!}\left\{
\frac{1}{4} 2^n (2 - h)^{-1 - n} \Gamma(1 + n) \right. \\
&\left. + \frac{2^{1 + n} (2 - h)^{-(3/2) - n}\Gamma(3/2 + n) {}_2F_1\left[1/2, 3/2 + n, 3/2, 
  1/(-2 + h)\right]}{2\sqrt{\pi}}\right. \\
& \left. + 1/4 \frac{(-1)^n 1}{\pi\sqrt{2}}\left.\frac{\partial^n}{\partial p^n}\left(\frac{1}{p\sqrt{1/2 + p}}\text{Arctan}\left(\frac{1}{\sqrt{1 + 2p}}\right) \right)\right|_{p=(1-h/2)} \right\},\\
l_{4} &=  \sqrt{2} \int_{-\infty}^{\infty} r\left(R\right) \left[10\Phi(R)^3 + 15\Phi(R)^2 + 12\Phi(R) - 6\right] \phi\left(\sqrt{2}R\right)dR\\
&=  12 l_2 - 6 l_1 + \frac{1}{g\sqrt{\pi}} 15 \sum_{n=0}^{\infty} \frac{(g)^{2n+1}}{(2n+1)!} \frac{(-1)^n }{\sqrt{2}\pi}\left.\frac{\partial^n}{\partial p^n}\left(\frac{1}{p\sqrt{1/2 + p}}\text{Arctan}\left(\frac{1}{\sqrt{1 + 2p}}\right) \right)\right|_{p=(1-h/2)}\\
&+\frac{45}{2g\sqrt{\pi}}\sum_{n=0}^{\infty} \frac{(g)^{2n+1}}{(2n+1)!} \frac{2^{1 + n} (2 - h)^{-(3/2) - n}\Gamma(3/2 + n) {}_2F_1\left[1/2, 3/2 + n, 3/2, 1/(-2 + h)\right]}{\sqrt{\pi}}\\
&+\frac{5}{g\sqrt{\pi}} \sum_{n=1}^{\infty} \frac{(g)^{n}}{n!} 2^{1/2 (-1 + n)} (1 + \exp(n \pi)) (2 - h)^{-(1/2) - n/2}\Gamma[(1 + n)/2]\\
&+ \frac{5}{4g\sqrt{\pi}} \sum_{n=1}^{\infty} \frac{(g)^{n}}{n!} \frac{1}{\pi (1-h/2)}\frac{3}{\sqrt{1 + 2(1-h/2)}}\text{Arctan}\left(\frac{1}{2\triangle\sqrt{1/2 + (1-h/2)}}\right),\\
& + O\left( \int_{-\infty}^{\infty} R^n \erf(R/\sqrt{2})^3 \exp\left(-(1-h/2)R^2\right)dR\right),
\end{aligned}
\end{equation}

with $\triangle = \sqrt{3/2 + (1-h/2)}$.
}
\end{proposition}

Then, given the L-Moments, for instance in the case of the g-and-h subfamily, the estimates of $g$ and $h$ are simultaneously found by iteratively minimising the objective
\begin{equation}
\label{eq:mlmom_gnh}
(\tau_{3} - \hat{\tau}_{3})^{2} + (\tau_{4} - \hat{\tau}_{4})^{2},
\end{equation}
subject to $0 \le h < 1$, where $\hat{\tau}_{3} = \hat{l}_{3} / \hat{l}_{2}$ is the sample L-skewness and $\hat{\tau}_{4} = \hat{l}_{4} / \hat{l}_{2}$ is the sample L-kurtosis. The estimates of $a$ and $b$ can be obtained using the following properties of L-moments.
\begin{proposition}[L-moments of affine functions of random variables]
Let $\mathbb{L}_{k}(\cdot)$ denote the $k$-th L-moment operator. Consider random variables $X$ and $Y$ such that $Y = a + bX$, where $a$ and $b$ are constants. The first and second L-moments of $Y$ can be expressed as
\begin{equation}
\label{eq:lmom_affine}
\begin{aligned}
\mathbb{L}_{1}(Y) &= a + b\mathbb{L}_{1}(X), \\
\mathbb{L}_{2}(Y) &= b\mathbb{L}_{2}(X).
\end{aligned}
\end{equation}
\end{proposition}
Given the estimates of $g$ and $h$, using \eqref{eq:lmom_affine}, one can estimate the values of $b$ and $a$ by
\begin{equation}
\label{eq:mlmom_bna}
\begin{aligned}
b &= \hat{l}_{2} / l_{2}, \\
a &= \hat{l}_{1} - bl_{1}.
\end{aligned}
\end{equation}

An alternative L-moment based approach can also be considered, where we consider a special choice of base distribution for $W$ given by the logistic model. If one modifies the Tukey transform family in the g-and-h case as follows it is also possible to obtain a re-parameterized form which admits closed form expressions for the L-moments. This particular sub-family case is known as the L-moment Tukey transformation families and it was first developed by \cite{headrick2012logistic}. The choice of logistic distribution for $W$ means that it will take a density, distribution, and quantile functions given by
\begin{align}
\begin{split}
f(w) &= \frac{ \exp\left( -{(w - \mu)}/{s}\right) }{ s\left( 1 + \exp\left( - {(w - \mu)}/{s}\right) \right)^2 }, \\
F(w) &= \frac{ 1 }{ 1 + \exp\left( - {(w - \mu)}/{s}\right)},\\
Q_W(\alpha) &= \mu + s\ln\left( \frac{ \alpha }{ 1 - \alpha } \right),\;\; \alpha \in [0,1],
\end{split}
\end{align}
for all $w \in \mathbb{R}$, $\mu \in \mathbb{R}$, and $s \in \mathbb{R}^+$. The motivation for modifying the distribution transformed under the Tukey structure was related to the fact that inference on the parameters can then be performed more readily via L-moments and L-correlation. The resulting four basic classes of modified Tukey quantile function transformations are then given in Definition \ref{DefnLTukey}.

\begin{definition}[L-Moment Tukey Transforms] \label{DefnLTukey}
Let $W \sim Logistic(\mu=0, s=1)$ be a standard logistic distributed random variable. Then the loss random variable $X$ has severity distribution given by the L-moment Tukey family as follows:
\begin{enumerate}
\item{The $\gamma{-}\kappa$ Tukey family transformation is given by
\begin{align}
X = T_{\gamma,\kappa}(W) = \gamma^{-1}\left( \exp(\gamma W) - 1 \right)\exp(\kappa |W|).
\end{align}
This is the analog of the g-and-h Tukey transform for the logistic distribution case for $\gamma \neq 0$ and $\kappa \geq 0$;
}
\item{The $\kappa_L{-}\kappa_R$ Tukey family transformation is given by
\begin{align}
X = T_{\kappa_L,\kappa_R}(W) = \begin{cases}
W\exp\left( \kappa_L |W|\right), & W \leq 0\\[4pt]
W\exp\left( \kappa_R |W|\right), & W \geq 0.
\end{cases}
\end{align}
This is the analog of the double h--h Tukey transform for the logistic distribution
case for $\kappa_L \geq 0$, $\kappa_R \geq 0$, and $\kappa_L \neq \kappa_R$.
}
\end{enumerate}
\end{definition}
These modified transformations then allow one to obtain the population L-moments in terms of the parameters of the L-moment $\gamma-\kappa$ Tukey family as well as the asymmetric L-moment $\kappa_L-\kappa_R$ Tukey family, which can be matched to the sample-estimated L-moments and then utilized as a system of nonlinear equations to be solve numerically via root search for the resulting L-moment parameter estimates.

\begin{proposition}[L-Moment Estimators for the L-Moment ${\boldsymbol \gamma}{\boldsymbol -}{\boldsymbol \kappa}$ Tukey Family]\label{PropLMEST}
Consider a $\gamma$-and-$\kappa$ distributed random variable $X \sim F(\gamma,\kappa)$ and a sample of $n$ loss data points with order statistics $\left\{X_{(i,n)}\right\}_{i=1}^n$ that will be used to fit the $\gamma$-and-$\kappa$ distribution. Then under the restrictions that $\gamma + \kappa < 1$, $\kappa < 1$, and $1+\gamma > \kappa$, which allow the first two L-moments to be finite, one obtains the following two equations for the population's first two L-moments $\lambda_1$ and $\lambda_2$ given by:
\begin{align}
\begin{split}
\lambda_1 &= \frac{ (-\gamma - \kappa)h_1 + (\gamma - \kappa)h_2 + (-\gamma + \kappa)h_3 + 2 \kappa h_4 +(\gamma + \kappa)h_5 - 2\kappa h_6 }{ 2\gamma }  \\
\lambda_2 &= \frac{ 2\gamma - (\gamma + \kappa)^2 h_1 + (\gamma - \kappa)^2\left(h_1 - h_3\right) + (\gamma + \kappa)^2 h_5  }{ 2 \gamma},
\end{split}
\end{align}
where $h_1,h_2,\ldots,h_6$ are defined with respect to the Harmonic number functions with the following arguments according to
\begin{align}
\begin{split}
h_1 &= H\left[\frac{1}{2}(-1-\gamma-\kappa)\right],\;\;\;\; h_2 = H\left[\frac{1}{2}(-1+\gamma-\kappa)\right]\\[4pt]
h_3 &= H\left[\frac{1}{2}(\gamma-\kappa)\right], \;\;\;\;  h_4 = H\left[\frac{1}{2}(-1-\kappa)\right]\\[4pt]
h_5 &= H\left[-\frac{1}{2}(\gamma+\kappa)\right], \;\;\;\;  h_6 = H\left[-\frac{1}{2}\kappa\right],
\end{split}
\end{align}
with the harmonic number functions defined for any $x > 0$ by
\begin{align}
H[x]:= x\sum_{k=1}^{\infty}\frac{1}{k(x+k)}.
\end{align}
One can then estimate sample L-moments  that can be matched to the population moments to solve numerically for the parameters.
\end{proposition}

\begin{remark}
As noted by \cite{headrick2012logistic}, expressions are also developed for the population L-skewness $\tau_3$ and L-kurtosis $\tau_4$ should one wish to utilize these for L-moment matching parameter estimation.
\end{remark}

Analogously, the solutions for the first two population L-moments for the class of $\kappa_L{-}\kappa_R$ Tukey transformations were detailed by \cite{headrick2012logistic} and can be used to perform parameter estimation, as detailed in Proposition \ref{PropLMESTKK}.

\begin{proposition}[L-Moment Estimators for the L-Moment ${\boldsymbol \kappa_L}\,{\boldsymbol -}\,{\boldsymbol \kappa_R}$ Tukey Family]\label{PropLMESTKK}
Consider the asymmetric $\kappa_L$-and-$\kappa_R$ distributed random variable $X \sim F(\kappa_L,\kappa_R)$ and a sample of $n$ loss data points with order statistics $\left\{X_{(i,n}\right\}_{i=1}^n$ that will be used to fit the$\kappa_L$-and-$\kappa_R$ distribution. Then under the restrictions that $\kappa_L < 1$ and $\kappa_R < 1$, which allow the first two L-moments to be finite, one obtains the following two equations for the population's first two L-moments $\lambda_1$ and $\lambda_2$ given~by
\begin{align}
\begin{split}
\lambda_1 &= \frac{1}{4}\left[  2p_5 - 2p_6 - 2p_7 + 2p_8 - \kappa_L p_9 + \kappa_L p_{10} + \kappa_R p_{11} - \kappa_R p_{12} \right] \\[4pt]
\lambda_2 &= \frac{1}{4}\left[ 4 + \kappa_L\left( -4 p_5 + 4 p_6 + \kappa_L\left( p_9 - p_{10}\right) \right) + 4 + \kappa_R\left( -4 p_7 + 4 p_8 + \kappa_R\left( p_{11} - p_{12}\right) \right) \right],
\end{split}
\end{align}
where $p_5,p_6,\ldots,p_{12}$ are defined with respect to the polygamma functions with the following arguments according to
\begin{align}
\begin{split}
p_5 &= P\left[0,\frac{1}{2} - \frac{\kappa_L}{2}\right],\;\;\; p_6 = P\left[0, 1 - \frac{\kappa_L}{2}\right],\;\;\; p_7 = P\left[0,\frac{1}{2} - \frac{\kappa_R}{2}\right]\\
p_8 &= P\left[0, 1 - \frac{\kappa_R}{2}\right],\;\;\; p_9 = P\left[1,\frac{1}{2} - \frac{\kappa_L}{2}\right],\;\;\; p_{10} = P\left[1, 1 - \frac{\kappa_L}{2}\right]\\
p_{11} &= P\left[1,\frac{1}{2} - \frac{\kappa_R}{2}\right],\;\;\; p_{12} = P\left[1, 1 - \frac{\kappa_R}{2}\right],
\end{split}
\end{align}
with the polygamma functions defined by
\begin{align}
P[m,x]:= (-1)^{m+1}m! \sum_{k=0}^{\infty} \frac{ 1 }{ (x + k)^{m+1} }.
\end{align}
One can then estimate sample L-moments that can be matched to the population L-moments to solve numerically for the parameters.
\end{proposition}

\section{Simulation study: Comparison of Estimation Procedures for Tukey Elongation Family Claims Models}
\label{sec:sim_study}

To investigate the quality of the various parameter estimation methods, including the approach we have developed in this manuscript based on L-moments, in fitting the g-and-h distributions, we conduct a simulation study, focusing on bias, variability, and computational cost of each method. The methods under comparison are moment matching (MoM), maximum likelihood (ML), quantile matching (QM), and method of L-moments (MoLM). 

Independent samples are generated from two g-and-h distributions whose parameters are $(0, 1, 0.1, 0.1)'$ and $(0, 1, 0.5, 0.2)'$, with the second one being more skewed and heavier tailed than the first. Notice that the first four moments are finite for both distributions as $h < \frac{1}{4}$ in both cases, and thus the MoM estimator is not being disadvantaged by design. Independent observations from a g-and-h distribution are generated by first generating $u_{i}$ from an uniform distribution on the interval $(0, 1)$ for $i \in \{1 \ddd n\}$. A sample from the g-and-h distribution is then obtained by applying the transformation, $x_{i} = X(u_{i})$, given by \eqref{eq:gh_qtlfun}. We use the Mersenne Twister (MT) pseudo-random number generator \cite{MatsumotoNishimura1998} to generate the uniform observations. Samples of sizes $n = 50$, $n = 100$, and $n = 1000$ are considered. We generate 1000 samples of each sample size from each g-and-h distribution.

Summaries of the estimates by the studied methods are reported in Table~\ref{tab:simresult_gh}. For each parameter, we report the Monte Carlo mean (Mean), standard deviation (SD), and mean squared error (MSE). We also report the mean and standard deviation of the per-estimation-time in seconds (Time) of each method, which allows us to assess the feasibility and scalability of the method.

It can be seen that for both g-and-h distributions, all sample sizes, and all four parameters, the MoLM has either lowest or equally lowest MSE amongst all considered estimators. The MoM has the highest MSE in most cases. As the sample size increases, the MSE is reduced for all estimators of all parameters, except for the MoM of the $h$ parameter. Comparing the Monte Carlo means, the MoM is particularly poor at estimating the $h$ parameter; the value of $h$ is significantly underestimated even for $n = 1000$. The MoLM slightly underestimates the $g$ and the $h$ parameters, however the bias is reducing with an increasing sample size. Comparing between the QM and the MoLM for the $g$ and the $h$ parameters, the mean of QM estimates is closer to the true value, however the MoLM has a lower standard deviation, especially for $n = 50$ and $n = 100$. 

Comparing the mean per-estimation-time amongst the estimators, the MoLM is the fastest for the first g-and-h distribution, and is sometimes slightly slower than MoM for the second one. The ML estimator is the slowest in each case and scales poorly with sample size. 

The conclusion of the results in this study demonstrate that the estimation of the g-and-h family of claims severity models is accurately and efficiently estimated using the proposed approach we have developed based around L-moments based estimation which was based on equating our derived population L-moments to the sample estimated L-moments and solving this system of non-linear equations. Importantly, the method was accurate for the h-parameter which dictates the kurtosis of the resulting claims severity distribution and will be important to be accurately and robustly estimated in practical settings. Furthermore, our proposed L-moments based approach was significantly more efficient computationally compared to other procedures, especially with increasing sample size.

\begin{table}[htbp]
\small
\centering
\begin{tabular}{ccc|rrrr|rrrr}
\toprule
&& $(a, b, g, h)$ & \multicolumn{4}{c}{(0, 1, 0.1, 0.1)} & \multicolumn{4}{|c}{(0, 1, 0.5, 0.2)} \\
          &       &       & MoM   & ML    & QM    & MoLM  & MoM   & ML    & QM    & MoLM \\
\midrule[\heavyrulewidth]
\multirow{14}{*}{$n = 50$} && Mean  & 0.002 & -0.002 & -0.001 & -0.001 & 0.001 & 0.007 & 0.005 & -0.001 \\
          & $a$   & SD    & 0.167 & 0.160 & 0.160 & 0.159 & 0.222 & 0.180 & 0.165 & 0.163 \\
          &       & MSE   & 0.028 & 0.026 & 0.026 & 0.025 & 0.049 & 0.032 & 0.027 & 0.027 \\
\cmidrule{2-11}
          &       & Mean  & 1.104 & 1.008 & 0.994 & 0.994 & 1.446 & 1.024 & 1.007 & 1.007 \\
          & $b$   & SD    & 0.142 & 0.151 & 0.153 & 0.148 & 0.382 & 0.190 & 0.208 & 0.193 \\
          &       & MSE   & 0.031 & 0.023 & 0.023 & 0.022 & 0.345 & 0.037 & 0.043 & 0.037 \\
\cmidrule{2-11}
          &       & Mean  & 0.090 & 0.100 & 0.100 & 0.095 & 0.425 & 0.470 & 0.480 & 0.463 \\
          & $g$   & SD    & 0.176 & 0.166 & 0.175 & 0.156 & 0.217 & 0.208 & 0.222 & 0.204 \\
          &       & MSE   & 0.031 & 0.028 & 0.031 & 0.024 & 0.053 & 0.044 & 0.050 & 0.043 \\
\cmidrule{2-11}
          &       & Mean  & 0.022 & 0.075 & 0.101 & 0.096 & 0.008 & 0.159 & 0.195 & 0.173 \\
          & $h$   & SD    & 0.028 & 0.086 & 0.103 & 0.081 & 0.023 & 0.122 & 0.170 & 0.111 \\
          &       & MSE   & 0.007 & 0.008 & 0.011 & 0.007 & 0.037 & 0.017 & 0.029 & 0.013 \\
\cmidrule{2-11}
          & Time  & Mean  & 0.177 & 18.749 & 0.607 & 0.142 & 0.142 & 21.192 & 0.631 & 0.193 \\
          & (sec) & SD    & 0.104 & 3.229 & 0.040 & 0.027 & 0.038 & 5.309 & 0.057 & 0.051 \\
\midrule[\heavyrulewidth]
\multirow{14}{*}{$n = 100$} && Mean  & 0.001 & 0.003 & 0.003 & 0.003 & -0.051 & 0.024 & 0.008 & 0.006 \\
          & $a$   & SD    & 0.124 & 0.112 & 0.113 & 0.112 & 0.194 & 0.343 & 0.114 & 0.114 \\
          &       & MSE   & 0.015 & 0.013 & 0.013 & 0.013 & 0.040 & 0.119 & 0.013 & 0.013 \\
\cmidrule{2-11}
          &       & Mean  & 1.082 & 1.007 & 0.996 & 0.996 & 1.413 & 1.032 & 1.015 & 1.017 \\
          & $b$   & SD    & 0.105 & 0.113 & 0.117 & 0.112 & 0.307 & 0.226 & 0.157 & 0.144 \\
          &       & MSE   & 0.018 & 0.013 & 0.014 & 0.013 & 0.264 & 0.052 & 0.025 & 0.021 \\
\cmidrule{2-11}
          &       & Mean  & 0.102 & 0.099 & 0.102 & 0.098 & 0.529 & 0.489 & 0.502 & 0.487 \\
          & $g$   & SD    & 0.153 & 0.115 & 0.123 & 0.113 & 0.207 & 0.171 & 0.170 & 0.157 \\
          &       & MSE   & 0.023 & 0.013 & 0.015 & 0.013 & 0.044 & 0.029 & 0.029 & 0.025 \\
\cmidrule{2-11}
          &       & Mean  & 0.040 & 0.085 & 0.100 & 0.098 & 0.008 & 0.175 & 0.192 & 0.178 \\
          & $h$   & SD    & 0.033 & 0.065 & 0.076 & 0.061 & 0.023 & 0.087 & 0.127 & 0.086 \\
          &       & MSE   & 0.005 & 0.004 & 0.006 & 0.004 & 0.037 & 0.008 & 0.016 & 0.008 \\
\cmidrule{2-11}
          & Time  & Mean  & 0.175 & 38.064 & 0.600 & 0.141 & 0.155 & 50.968 & 0.616 & 0.192 \\
          & (sec) & SD    & 0.107 & 5.837 & 0.046 & 0.021 & 0.043 & 188.407 & 0.044 & 0.044 \\
\midrule[\heavyrulewidth]
\multirow{14}{*}{$n = 1000$} && Mean  & -0.001 & 0.001 & 0.001 & 0.001 & -0.180 & -0.185 & 0.001 & 0.001 \\
          & $a$   & SD    & 0.043 & 0.037 & 0.036 & 0.036 & 0.107 & 2.590 & 0.036 & 0.036 \\
          &       & MSE   & 0.002 & 0.001 & 0.001 & 0.001 & 0.044 & 6.744 & 0.001 & 0.001 \\
\cmidrule{2-11}
          &       & Mean  & 1.028 & 1.002 & 0.999 & 0.999 & 1.188 & 1.424 & 1.003 & 1.003 \\
          & $b$   & SD    & 0.042 & 0.038 & 0.036 & 0.035 & 0.130 & 3.606 & 0.049 & 0.045 \\
          &       & MSE   & 0.003 & 0.001 & 0.001 & 0.001 & 0.052 & 13.183 & 0.002 & 0.002 \\
\cmidrule{2-11}
          &       & Mean  & 0.103 & 0.099 & 0.099 & 0.100 & 0.791 & 0.474 & 0.502 & 0.500 \\
          & $g$   & SD    & 0.058 & 0.036 & 0.038 & 0.036 & 0.142 & 0.190 & 0.053 & 0.052 \\
          &       & MSE   & 0.003 & 0.001 & 0.001 & 0.001 & 0.105 & 0.037 & 0.003 & 0.003 \\
\cmidrule{2-11}
          &       & Mean  & 0.082 & 0.098 & 0.100 & 0.099 & 0.005 & 0.182 & 0.199 & 0.197 \\
          & $h$   & SD    & 0.022 & 0.021 & 0.024 & 0.020 & 0.015 & 0.055 & 0.046 & 0.031 \\
          &       & MSE   & 0.001 & 0.000 & 0.001 & 0.000 & 0.038 & 0.003 & 0.002 & 0.001 \\
\cmidrule{2-11}
          & Time  & Mean  & 0.152 & 390.220 & 0.607 & 0.130 & 0.267 & 539.302 & 0.599 & 0.188 \\
          & (sec) & SD    & 0.090 & 76.362 & 0.057 & 0.021 & 0.132 & 206.724 & 0.036 & 0.039 \\
\bottomrule
\end{tabular}
\caption{Summaries of g-and-h parameter estimates by various methods on simulated data.}
\label{tab:simresult_gh}
\end{table}
\FloatBarrier

\section{Real Data Empirical Application}
\label{sec:emp_study}
In this section we consider a real data study involving a very large data set of claims records from Australia. In this study we consider fitting the g, h, g-and-h, and g-and-k distributions which are employed to model the total gross payment of individual claims of the Compulsory Third Party (CTP) insurance from an insurance company based in Queensland, Australia. The data contains 115,300 accident records from September 1994 to December 2008. Unlike most models in non-life insurance where the annual aggregate amounts of claims are modelled, in this paper, we are interested in modelling the distributions of individual claim payment amounts, not the aggregate for a claims development year. In this way we resolve to a better degree the actual claims distribution, which we can achieve through use of a flexible severity loss model. Two key challenges of modelling such data are: the computational burden associated with the large number of observations, and the models for aggregate amounts may not be flexible enough to adequately model the losses at the claim level. 

In undertaking this study, we are also interested in studying the time varying skewness and kurtosis of the claims records over time, to achieve this we consider to split the records into monthly time series and study the distributional properties each month, allowing us to study temporal variation in skewness and kurtosis of such claims annualy.

\begin{table}[ht]
\small
\centering
\begin{tabular}{ccccccc}
\toprule
        &  $a$  &  $b$  &  $g$   &  $h$  &  $k$  &  RMSE \\
\midrule
g       & 9.571 & 1.794 & -0.238 &   -   &   -   & 0.143 \\
h       & 9.355 & 1.717 &   -    & 0.053 &   -   & 0.307 \\
g-and-h & 9.566 & 1.717 & -0.230 & 0.035 &   -   & 0.179 \\
g-and-k & 9.568 & 1.665 & -0.293 &   -   & 0.088 & 0.138 \\
\bottomrule
\end{tabular}
\caption{MoLM parameter estimates using all claims.}
\label{tab:est_all}
\end{table}
\FloatBarrier

\begin{figure}[ht]
\begin{center}
\includegraphics[width=0.8\textwidth]{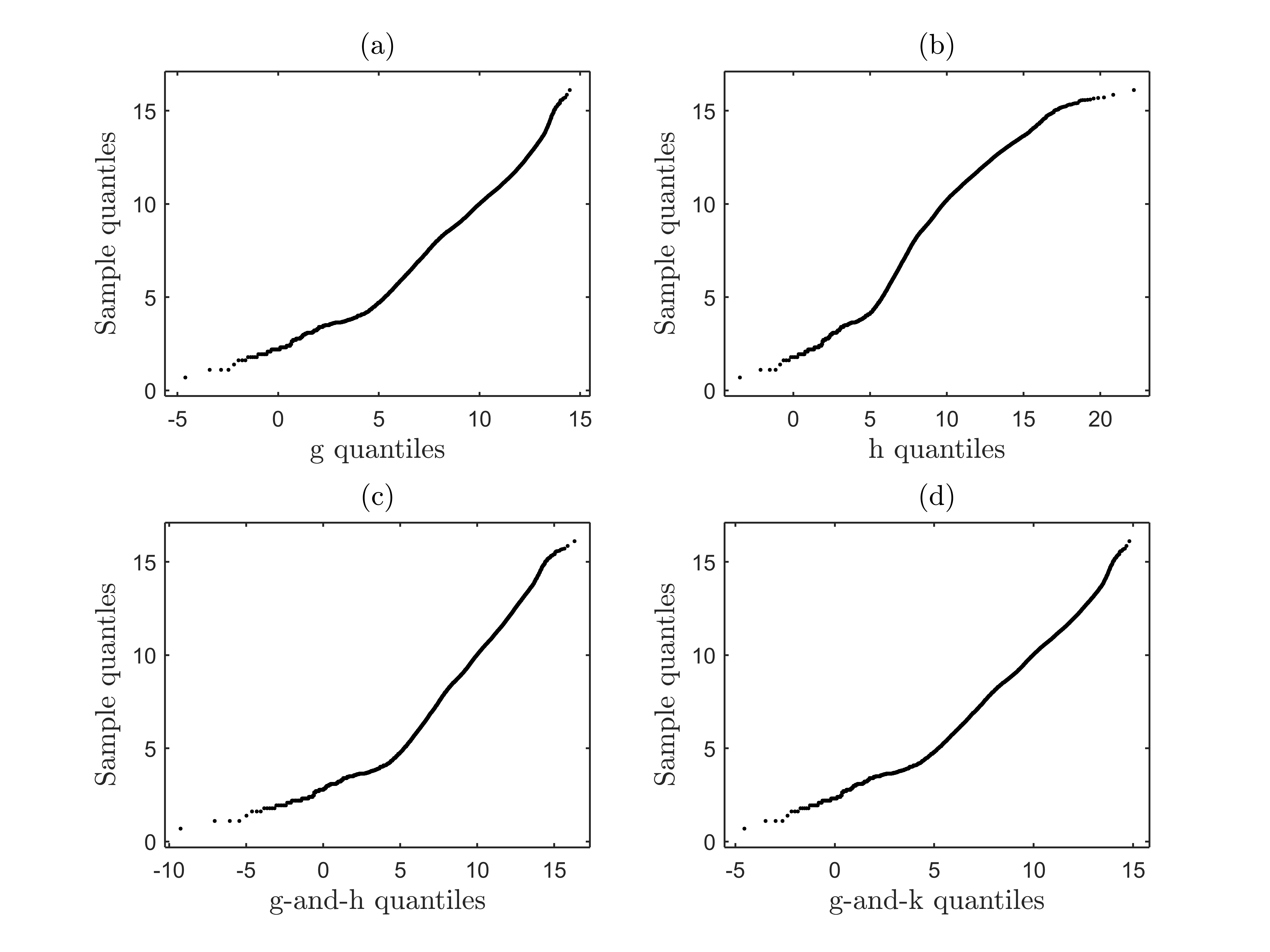}
\end{center}
\caption{Q-Q plot sample quantiles vs. model quantiles.}
\label{fig:qq_full}
\end{figure}

Table~\ref{tab:est_all} reports the parameters of the g, h, g-and-h, and g-and-k distributions fitted to the log total gross payment of all the claims in the sample. All models are estimated using the method of L-moment. To compare the goodness-of-fit between the models, we compute the root-mean-square-error (RMSE) between the model quantiles and the sample quantiles. Specifically
\begin{equation}
\label{eq:rmse}
\mathrm{RMSE} = \left[\frac{1}{n}\sum_{1}^{n}[X(u_{i}) - \hat{\chi}_{u_{i}}]^{2}\right]^{1/2},
\end{equation}
where $n$ is the number of observations, $X(u_{i})$ is the model quantile, $\hat{\chi}_{u_{i}}$ is the sample quantile, and the quantile level $u_{i}$ is given by
\begin{equation*}
u_{i} = \frac{i - 0.5}{n}.
\end{equation*}
The estimates of the $g$ indicates that the log payments are negatively skewed. From the estimates of $h$ and $k$, we see that the fitted distributions have tails that are only slightly heavier than the Gaussian tails. The values of the RMSE show that, for the entire sample, the g-and-k distribution provides the best fit, followed by the g distribution, while the h distribution fits the data the least well. The fact that the g and the g-and-k distributions show a similar fit indicates that for this type of data it tends to be more important to model the asymmetry than the heavy-tailedness. A Q-Q plot of the sample quantiles vs. the quantiles of the fitted model is shown for each of the models in Figure~\ref{fig:qq_full}.

\FloatBarrier

\begin{figure}[ht]
\begin{center}
\includegraphics[width=0.8\textwidth]{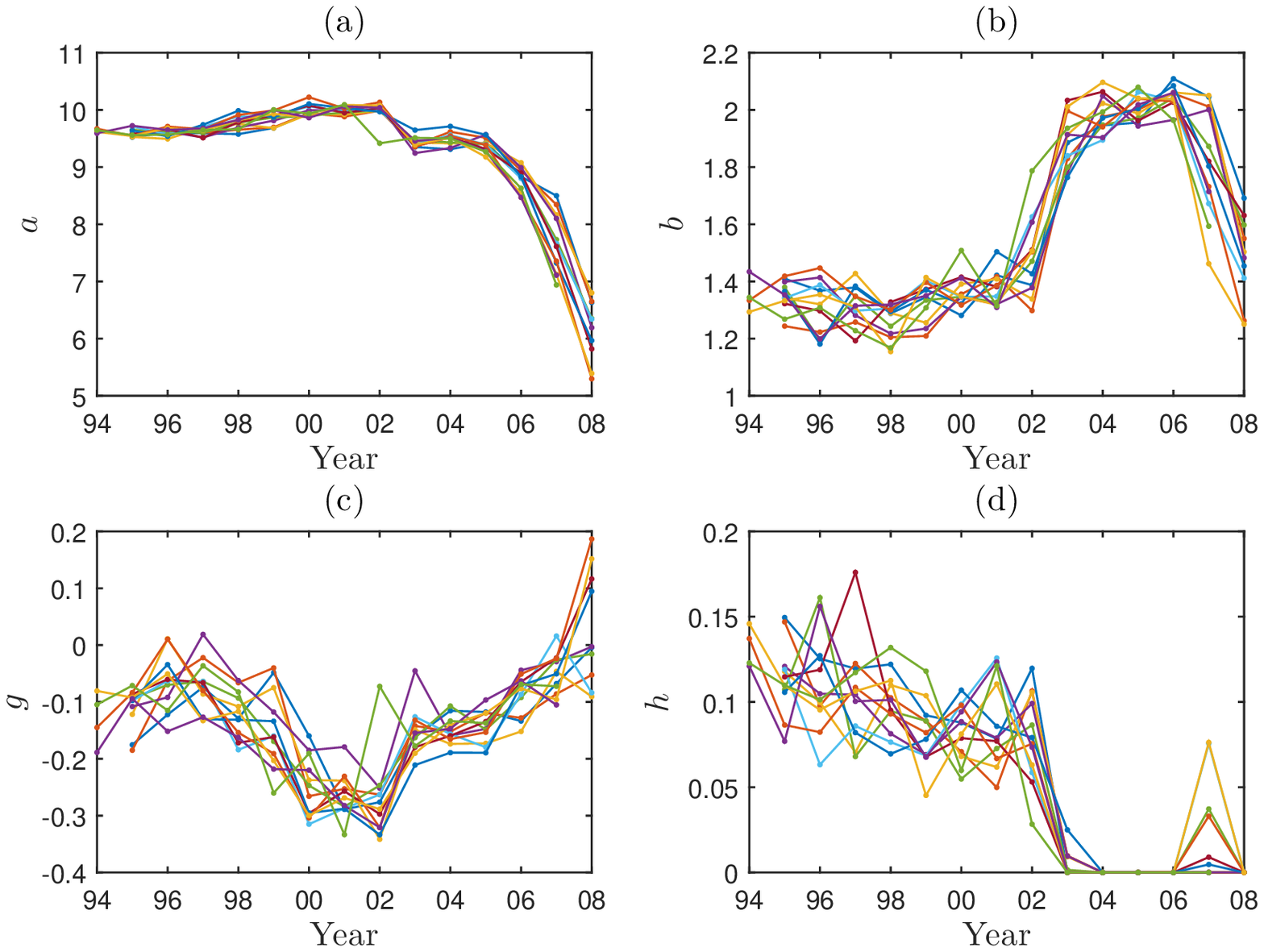}
\end{center}
\caption{}
\label{fig:est_gh_monthly}
\end{figure}
\FloatBarrier

\begin{figure}[ht]
\begin{center}
\includegraphics[width=0.8\textwidth]{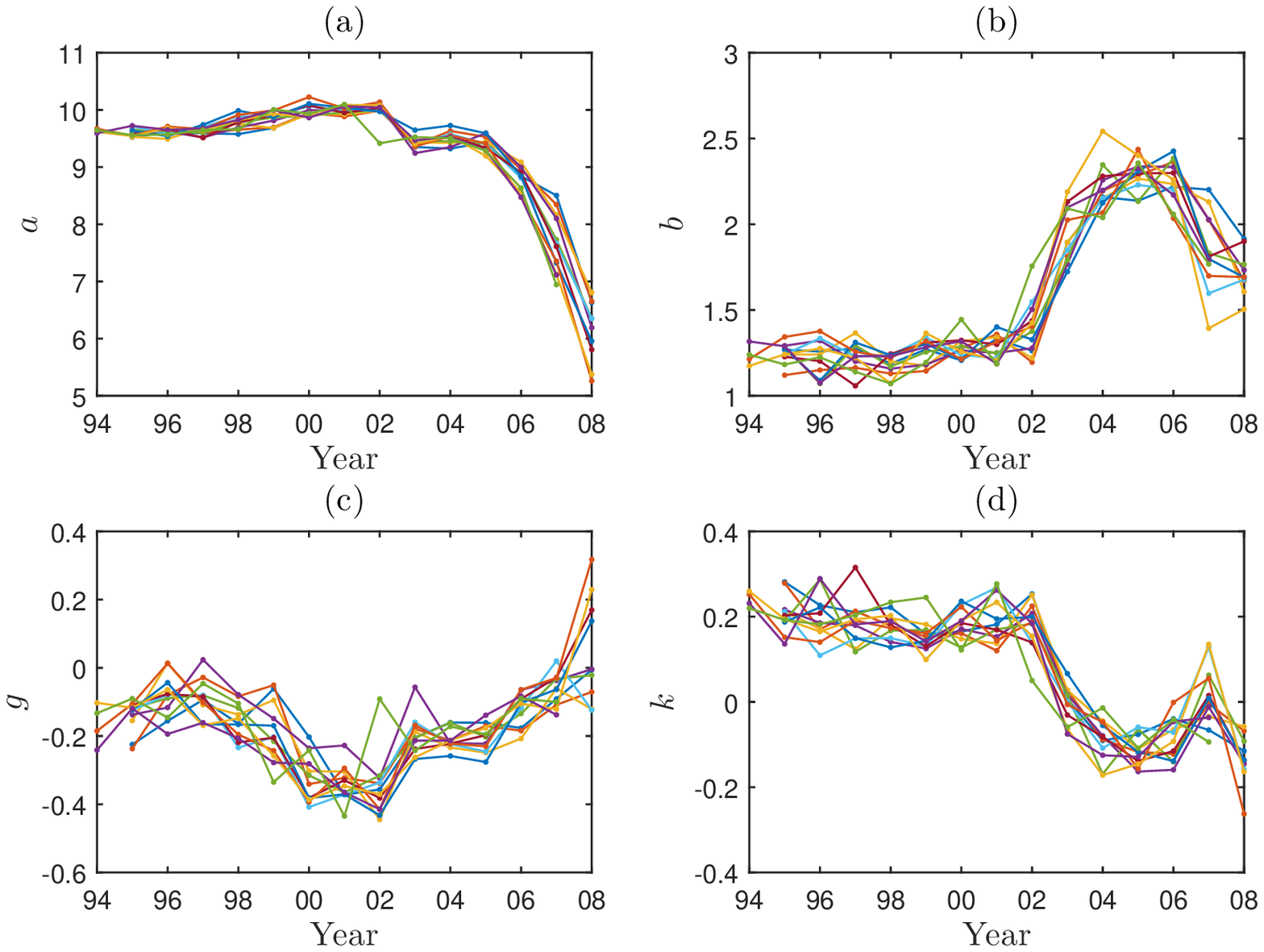}
\end{center}
\caption{}
\label{fig:est_gk_monthly}
\end{figure}
\FloatBarrier

We are also interested in how the parameters of the g-and-h and g-and-k distributions evolve over time. For this purpose, we estimate the two models using log payments from each month of the year. In Figure~\ref{fig:est_gh_monthly}, each line is created by plotting the parameter estimate of a specific month over the years covered by the sample (1994 to 2008). A similar plot for the g-and-k distribution is shown in Figure~\ref{fig:est_gh_monthly}. It seems that there was a structural change after 2002, where both distributions increased in scale but became less heavy-tailed. The locations of the distributions also started to shift downward.

\begin{figure}[ht]
\begin{center}
\includegraphics[width=0.7\textwidth]{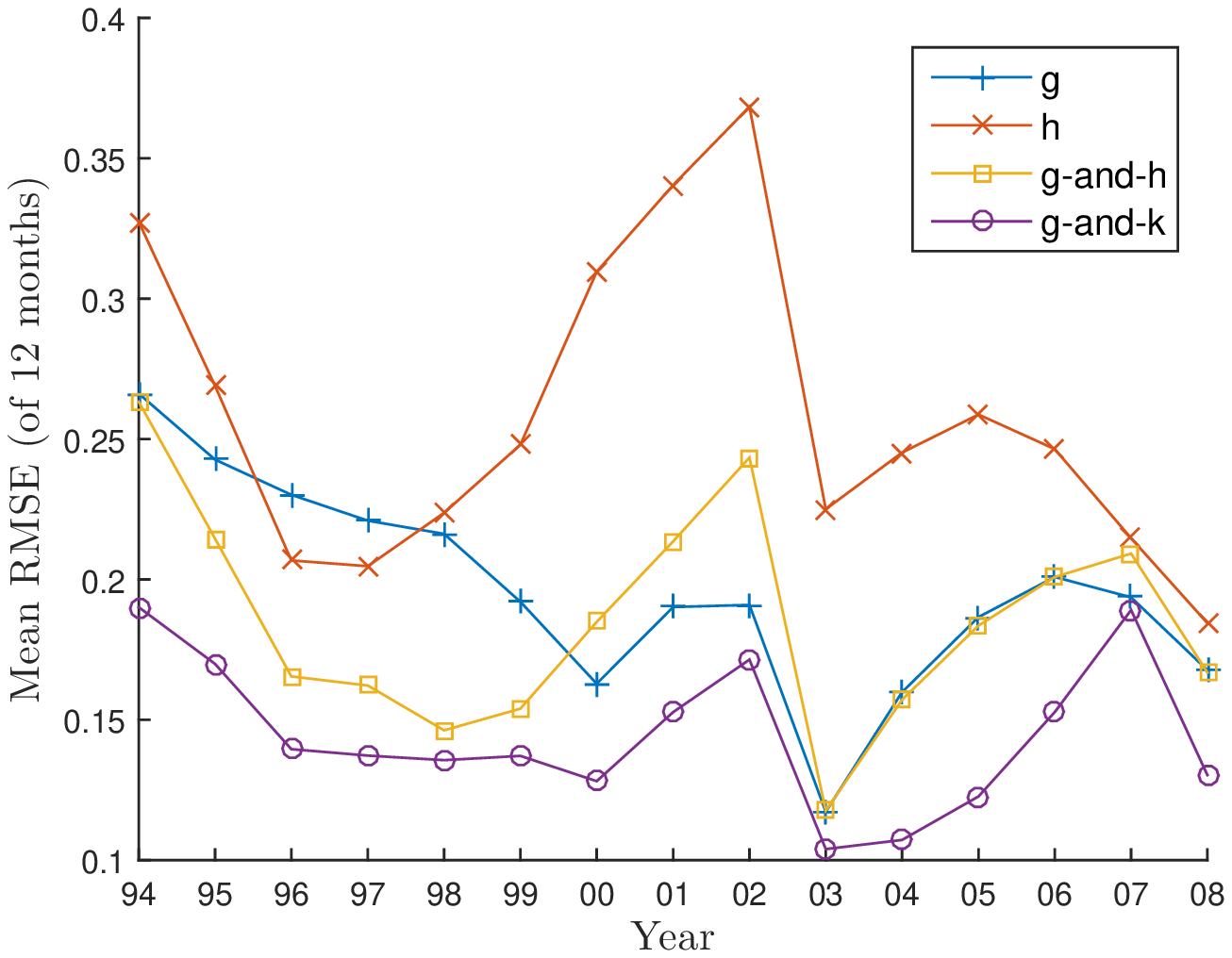}
\end{center}
\caption{}
\label{fig:mean_rmse}
\end{figure}
\FloatBarrier

To compare the goodness-of-fit between the models for each year, we compute RMSE in Equation \eqref{eq:rmse} for each month of the year and then plot the average of the 12 monthly RMSE values for each year in the sample. The plot of the average RMSE values for each model in shown Figure~\ref{fig:mean_rmse}. Once again, the g-and-k distribution provides the best fit in every year and the h distribution is worst for most of the years.

\section{Conclusion}

In this paper we have proposed a new family of claims reserving models for non-life insurance severity modelling corresponding to a flexible class of Tukey elongation transform models. We have outlined the characterization of the sub-families of g, h, k, j, h-h, g-and-h, g-and-k, g-and-j and generalized g variants. Furthermore, we have studied the properties of such models, including deriving novel relationships for the tail behaviour and the population L-moments for such models. Furthermore, we have demonstrated the estimation of these claims model parameters can accurately and efficiently be performed using L-method of moment matching, which can often be more accurate and efficient that alternative procedures based on moment-matching or maximum likelihood. Finally, we have applied the methods and models to calibration of claims severity models for a large insurance database based on CTP automotive claims data from Australia.

\bibliographystyle{apalike}
\bibliography{ghlmom}

\appendix

\section{Appendix A}

The following integral identities from \cite{prudnikov1998integrals} are useful for the following derivations of the population L-moments.

{\small{
\begin{equation}
\begin{split}
&\Phi(x) = \frac{1}{2}\text{erfc}\left(-\frac{x}{\sqrt{2}}\right)\\
&\int_{-\infty}^{\infty}\Phi(x)dx = \sqrt{2\pi}\\
&\int_{-\infty}^{\infty}\Phi(x)^2\phi(x)dx = \frac{1}{\pi}\text{Arctan}\left(\sqrt{3}\right)\\
&\int_{-\infty}^{\infty}\exp(cx - x^2/2) dx = \sqrt{2\pi}\exp\left(\frac{c^2}{2}\right)\\
&\int_{-\infty}^{0}\exp(cx)\phi(bx)^n dx = \frac{\exp\left(\frac{c^2}{2nb^2}\right)}{b\sqrt{n(2\pi)^{n-1}}}\Phi\left(\frac{b^2 x n - c}{b\sqrt{n}}\right) + C\\
&\int_{-\infty}^{0}\Phi(bx) \phi\left(ax\right)dx = \frac{1}{2\pi|a|}\left(\frac{\pi}{2} - \text{arctan}\left(\frac{b}{|a|}\right)\right)\\
&\int_{0}^{\infty}\Phi(bx) \phi\left(ax\right)dx =\frac{1}{2\pi|a|}\left(\frac{\pi}{2} + \text{arctan}\left(\frac{b}{|a|}\right)\right) \\
&\int_{0}^{\infty} \text{erfc}(cx)\exp\left(-px^2\right)x^n dx = I_n, \;\; \mathcal{R}\left\{c^2+p\right\}>0\\
&\;\;\;\;I_1 = \frac{1}{2p}\left(1 - \frac{c}{\sqrt{c^2 + p}}\right),\\
&\;\;\;\;I_{2m} = \frac{(-1)^m}{\sqrt{\pi}}\frac{\partial^m}{\partial p^m}\left(\frac{1}{\sqrt{p}}\text{Arctan}\left(\frac{\sqrt{p}}{c}\right) \right) \\
&\;\;\;\;I_{2m+1} = \frac{(-1)^m m!}{2p^{m+1}} - \frac{(-1)^m}{2}\frac{\partial^m}{\partial p^m}\left(\frac{1}{p\sqrt{p+c^2}}\right)\\
&\int_{0}^{\infty}x^{a-1}\text{erf}(cx)\exp\left(-px - bx^2\right) dx = \frac{c}{\sqrt{\pi}b^{(a+1)/2}}\Gamma\left(\frac{a+1}{2}\right)\psi_1\left(\frac{a+1}{2}, \frac{1}{2}, \frac{3}{2}, \frac{1}{2}; - \frac{c^2}{b}, \frac{p^2}{4b} \right)\\
&\;\;\;\;- \frac{cp}{\sqrt{\pi}b^{a/2 + 1}}\Gamma\left(\frac{a}{2}+1\right)\psi_1\left(\frac{a}{2}+1, \frac{1}{2}, \frac{3}{2}, \frac{3}{2}; - \frac{c^2}{b}, \frac{p^2}{4b} \right)
\\ 
\end{split}
\end{equation}
}}

{\small{
\begin{equation}
\begin{split}
&\int_{0}^{\infty}\text{erf}^2(cx)\exp\left(-px^2\right)x^{2n+1} dx = \frac{(-1)^n c}{\pi}\frac{\partial^n}{\partial p^n}\left(\frac{1}{p\sqrt{c^2 + p}}\text{Arctan}\left(\frac{c}{\sqrt{c^2 + p}}\right) \right), \;\; \mathcal{R}\left\{p > 0\right\}\\
&\int_{0}^{\infty}\text{erf}(ax)\text{erf}(bx)\text{erf}(cx)\exp\left(-px^2\right)x dx \\
& = \frac{1}{\pi p} \left[\frac{a}{\sqrt{a^2 + p}}\text{Arctan}\left(\frac{bc}{\triangle\sqrt{a^2 + p}}\right) + \frac{b}{\sqrt{b^2 + p}}\text{Arctan}\left(\frac{ac}{\triangle\sqrt{b^2 + p}}\right) + \frac{c}{\sqrt{c^2 + p}}\text{Arctan}\left(\frac{ab}{\triangle\sqrt{c^2 + p}}\right) \right],\\
&\triangle = \sqrt{a^2 + b^2 + c^2 + p}, \;\; \mathcal{R}\left\{p > 0\right\}\\
\end{split}
\end{equation}
}}

\subsection{Proof of Proposition \ref{PropG}}

\begin{proof}
It will also be useful to recall the integral identities from \cite{prudnikov1998integrals} which are given in Appendix A which are used to make these proofs for each of the first four L-moments as follows:
\begin{equation} 
\begin{aligned}
l_{1} &=  \sqrt{2} \int_{-\infty}^{\infty} r\left(R\right) \phi\left(\sqrt{2}R\right)dR =\frac{\sqrt{2}}{g} \int_{-\infty}^{\infty} \left(\exp(gR)-1\right) \phi\left(\sqrt{2}R\right)dR =\frac{\exp\left(\frac{g^2}{2}\right)-1}{g},\\
\end{aligned}
\end{equation}

\begin{equation} 
\begin{aligned}
l_{2} &=  \sqrt{2} \int_{-\infty}^{\infty} r\left(R\right)\Phi(R) \phi\left(\sqrt{2}R\right)dR\\
&=\frac{\sqrt{2}}{2g} \int_{0}^{\infty}\exp\left(gR - R^2\right)\text{erfc}\left(-\frac{R}{\sqrt{2}}\right)dR +\frac{\sqrt{2}}{2g} \int_{0}^{\infty}\exp\left(-gR - R^2\right)\text{erfc}\left(-\frac{R}{\sqrt{2}}\right)dR\\
&\;\;\;\; -\frac{\sqrt{2}}{g} \int_{-\infty}^{0}\Phi(R) \phi\left(\sqrt{2}R\right)dR  -\frac{\sqrt{2}}{g} \int_{0}^{\infty}\Phi(R) \phi\left(\sqrt{2}R\right)dR\\
%
%
&=\frac{\sqrt{2\pi}}{2g}\exp\left(g^2\right) - \frac{1}{2g} +\frac{2}{\sqrt{2\pi}}\psi_1\left(1, \frac{1}{2}, \frac{3}{2}, \frac{1}{2}; - \frac{1}{2}, \frac{g^2}{4} \right)\\
\end{aligned}
\end{equation}

\begin{equation} 
\begin{aligned}
l_{3} &=  \sqrt{2} \int_{-\infty}^{\infty} r\left(R\right) \left[3\Phi(R)^2 + 3\Phi(R) + 1\right] \phi\left(\sqrt{2}R\right)dR \\
&= 3\frac{\sqrt{2}}{g} \int_{-\infty}^{\infty} \exp\left(gR - R^2\right)\Phi(R)^2dR - \frac{3\text{Arctan}\left(\sqrt{3}\right)}{g\pi\sqrt{\pi}} + 3l_2 + l_1 \\
&= \frac{3\sqrt{2}}{4g} \left\{\int_{-\infty}^{\infty} \exp\left(gR - R^2\right)\text{erf}\left(\frac{R}{\sqrt{2}}\right)^2dR + 2 \int_{-\infty}^{\infty} \exp\left(gR - R^2\right)\text{erf}\left(\frac{R}{\sqrt{2}}\right)dR 
\right\}\\
&+ \frac{3\sqrt{2\pi}}{4g} \exp\left(\frac{g^2}{4}\right) - \frac{3\text{Arctan}\left(\sqrt{3}\right)}{g\pi\sqrt{\pi}} + 3l_2 + l_1 \\
&= \frac{3}{g\sqrt{\pi}} \psi_1\left(1, \frac{1}{2}, \frac{3}{2}, \frac{1}{2}; - \frac{1}{2}, \frac{g^2}{4} \right) + \frac{3\sqrt{2\pi}}{4g} \exp\left(\frac{g^2}{4}\right) - \frac{3\text{Arctan}\left(\sqrt{3}\right)}{g\pi\sqrt{\pi}} + 3l_2 + l_1 \\
&+ \frac{3\sqrt{2}}{4g} \left\{\sum_{n=0}^{\infty}\int_{0}^{\infty} \frac{(gR)^n}{n!}\exp\left(-R^2\right)\text{erf}\left(\frac{R}{\sqrt{2}}\right)^2dR + \int_{0}^{\infty}\frac{(gR)^n}{n!}\exp\left(-gR - R^2\right)\text{erf}\left(\frac{R}{\sqrt{2}}\right)^2dR \right\}\\
%
&= \frac{3}{g\sqrt{\pi}} \psi_1\left(1, \frac{1}{2}, \frac{3}{2}, \frac{1}{2}; - \frac{1}{2}, \frac{g^2}{4} \right) + \frac{3\sqrt{2\pi}}{4g} \exp\left(\frac{g^2}{4}\right) - \frac{3\text{Arctan}\left(\sqrt{3}\right)}{g\pi\sqrt{\pi}} + 3l_2 + l_1 \\
\end{aligned}
\end{equation}

\begin{equation} 
\begin{aligned}
l_{4} &=  \sqrt{2} \int_{-\infty}^{\infty} r\left(R\right) \left[10\Phi(R)^3 + 15\Phi(R)^2 + 12\Phi(R) - 6\right] \phi\left(\sqrt{2}R\right)dR\\
&= 10\sqrt{2} \int_{-\infty}^{\infty} r\left(R\right)\Phi(R)^3 \phi\left(\sqrt{2}R\right)dR + 5 l_3 + \frac{45\text{Arctan}\left(\sqrt{3}\right)}{g\pi\sqrt{\pi}} + 27 l_2 - 1 l_1\\
&= 5 l_3 + \frac{45\text{Arctan}\left(\sqrt{3}\right)}{g\pi\sqrt{\pi}} + 27 l_2 - 1 l_1 + \frac{20\sqrt{2}}{g} \sum_{n=1}^{\infty}\frac{(g)^{2n+1}}{(2n+1)!} \int_{0}^{\infty} R^{2n+1}\Phi(R)^3 \phi\left(\sqrt{2}R\right)dR\\
&= 5 l_3 + \frac{45\text{Arctan}\left(\sqrt{3}\right)}{g\pi\sqrt{\pi}} + 27 l_2 - 1 l_1 + \frac{30\sqrt{2}}{\sqrt{3}\pi}\text{Arctan}\left(\frac{1}{\sqrt{15}}\right) + O\left(\int_{0}^{\infty} R^{3}\Phi(R)^3 \phi\left(\sqrt{2}R\right)dR \right)
\end{aligned}
\end{equation}
\end{proof}

\subsection{Proof of Proposition \ref{PropH}}

\begin{proof}
The first four population L-Moments of the h-family of Tukey transform Loss models is given for $a=0$, $b=1$ by considering $$r(R) = R\exp\left(\frac{hR^2}{2}\right)$$ giving:
\begin{equation}
\begin{aligned}
l_{1} &=  \sqrt{2} \int_{-\infty}^{\infty} r\left(R\right) \phi\left(\sqrt{2}R\right)dR\\
&=  \frac{1}{\sqrt{\pi}} \int_{-\infty}^{\infty} R \exp\left(-(1-h)R^2\right)dR = 0, \;\; h < 1\\
\end{aligned}
\end{equation}

\begin{equation}
\begin{aligned}
l_{2} &=  \sqrt{2} \int_{-\infty}^{\infty} r\left(R\right)\Phi(R) \phi\left(\sqrt{2}R\right)dR\\
&= \frac{1}{\sqrt{\pi}} \int_{-\infty}^{\infty} R \Phi(R) \exp\left(-(1-h)R^2\right)dR\\
&= \frac{1}{\sqrt{\pi}} \int_{0}^{\infty} R \text{erfc}\left(-\frac{R}{\sqrt{2}}\right) \exp\left(-(1-h)R^2\right)dR\\
&= \frac{1}{\sqrt{\pi}} \frac{1}{(1-h)}\left(1 + \frac{1}{\sqrt{1+2(1-h)}}\right)\\
\end{aligned}
\end{equation}

\begin{equation}
\begin{aligned}
l_{3} &=  \sqrt{2} \int_{-\infty}^{\infty} r\left(R\right) \left[3\Phi(R)^2 + 3\Phi(R) + 1\right] \phi\left(\sqrt{2}R\right)dR \\
&=  \frac{3}{\sqrt{\pi}} \int_{-\infty}^{\infty} R\text{erfc}^2\left(-\frac{R}{\sqrt{2}}\right) \exp\left(-(1-h)R^2\right)dR + 3 l_2 + l_1\\
&= 3 l_2 + l_1 + \frac{3}{\sqrt{\pi}} \left\{
\int_{-\infty}^{\infty} R\exp\left(-(1-h)R^2\right)dR
-2 \int_{-\infty}^{\infty} R\text{erf}\left(-\frac{R}{\sqrt{2}}\right)\exp\left(-(1-h)R^2\right)dR 
\right\}\\
& + \frac{6}{\sqrt{\pi}}  \int_{-\infty}^{\infty} R\text{erf}^2\left(-\frac{R}{\sqrt{2}}\right)\exp\left(-(1-h)R^2\right)dR \\
&= 3 l_2 + l_1 + \frac{6}{\sqrt{\pi}}  \int_{-\infty}^{\infty} R\text{erf}^2\left(-\frac{R}{\sqrt{2}}\right)\exp\left(-(1-h)R^2\right)dR \\
&= 3 l_2 + l_1 \\
\end{aligned}
\end{equation}

\begin{equation}
\begin{aligned}
l_{4} &=  \sqrt{2} \int_{-\infty}^{\infty} r\left(R\right) \left[10\Phi(R)^3 + 15\Phi(R)^2 + 12\Phi(R) - 6\right] \phi\left(\sqrt{2}R\right)dR\\
&=  12 l_2 - 6 l_1 + \frac{10}{\sqrt{\pi}} \int_{-\infty}^{\infty} R \left[1/2 + 1/2 \text{erf}\left(\frac{R}{sqrt{2}}\right)\right]^3 \exp\left(-(1-h)R^2\right)dR\\
&= \frac{10}{8\sqrt{2\pi}} l_3 + 12 l_2 - 6 l_1 + 
\frac{10}{8\sqrt{\pi}} \int_{-\infty}^{\infty} R 
\text{erf}^3\left(\frac{R}{sqrt{2}}\right)\exp\left(-(1-h)R^2\right)dR\\
&= \frac{10}{8\sqrt{2\pi}} l_3 + 12 l_2 - 6 l_1 + 
\frac{30}{8\pi(1-h)\sqrt{\pi[1 + 2(1-h)]}} 
\text{Arctan}\left(\frac{1}{\sqrt{[2 + 4(1-h)][\frac{3}{2}+(1-h)]}}\right) 
\end{aligned}
\end{equation}

\end{proof}

\subsection{Proof of Proposition \ref{PropK}}

\begin{proof}
The first four population L-Moments of the k-family of Tukey transform Loss models is given for $a=0$, $b=1$ by considering $$r(R) = R\left(1+R^2\right)^k$$ giving:
\begin{equation}
\begin{aligned}
l_{1} &=  \sqrt{2} \int_{-\infty}^{\infty} r\left(R\right) \phi\left(\sqrt{2}R\right)dR\\
&=  \sqrt{2} \int_{-\infty}^{\infty} R\left(1+R^2\right)^k \phi\left(\sqrt{2}R\right)dR = 0 \\
l_{2} &=  \sqrt{2} \int_{-\infty}^{\infty} r\left(R\right)\Phi(R) \phi\left(\sqrt{2}R\right)dR\\
&=  \frac{1}{\sqrt{2\pi}} \int_{-\infty}^{\infty}  \left\{R+k R^3+1/2 (k-1) k R^5+1/6 (k-2) (k-1) k R^7 \right\} \text{erfc}\left(-\frac{R}{\sqrt{2}}\right) \exp\left(-R^2\right)dR \\ 
&\;\;\;\;+O\left(\int_{-\infty}^{\infty}R^9 \Phi(R) \exp\left(-R^2\right)dR\right)\\
&= \frac{1}{\sqrt{2\pi}} \left\{I_1 + \widetilde{I}_1 +2k I_3 + (k-1) k I_5 +1/3 (k-2) (k-1) k I_7 \right\} +O\left(\int_{-\infty}^{\infty}R^9 \Phi(R) \exp\left(-R^2\right)dR\right)\\
\end{aligned}
\end{equation}

\begin{equation}
\begin{aligned}
l_{3} &=  \sqrt{2} \int_{-\infty}^{\infty} r\left(R\right) \left[3\Phi(R)^2 + 3\Phi(R) + 1\right] \phi\left(\sqrt{2}R\right)dR \\
&=  3\sqrt{2} \int_{-\infty}^{\infty} r\left(R\right) \Phi(R)^2 \phi\left(\sqrt{2}R\right)dR 
+ 3 l_2 + l_1 +O\left(\int_{-\infty}^{\infty}R^9 \Phi(R) \exp\left(-R^2\right)dR\right)\\
&= \frac{3}{2\sqrt{2\pi}} \int_{0}^{\infty}  \left\{R+k R^3+1/2 (k-1) k R^5+1/6 (k-2) (k-1) k R^7 \right\} \text{erf}^2\left(\frac{R}{\sqrt{2}}\right) \exp\left(-R^2\right)dR \\ 
&+ \int_{0}^{\infty}  \left\{R+k R^3+1/2 (k-1) k R^5+1/6 (k-2) (k-1) k R^7 \right\} \text{erf}^2\left(-\frac{R}{\sqrt{2}}\right) \exp\left(-R^2\right)dR \\ 
&+ 3 l_2 + \left(\frac{3}{4\sqrt{\pi}} + 1\right)l_1 +O\left(\int_{-\infty}^{\infty}R^9 \Phi(R) \exp\left(-R^2\right)dR\right)\\
&= \frac{3}{2\sqrt{2\pi}} \left\{(G_1 + \widetilde{G}_1) + k (G_{3} + \widetilde{G}_3) + 1/2 (k-1) k (G_{5} + \widetilde{G}_5) + 1/6 (k-2) (k-1) k (G_{7}+\widetilde{G}_7) \right\} \\ 
&+ 3 l_2 + \left(\frac{3}{4\sqrt{\pi}} + 1\right)l_1 +O\left(\int_{-\infty}^{\infty}R^9 \Phi(R) \exp\left(-R^2\right)dR\right),\\
l_{4} &=  \sqrt{2} \int_{-\infty}^{\infty} r\left(R\right) \left[10\Phi(R)^3 + 15\Phi(R)^2 + 12\Phi(R) - 6\right] \phi\left(\sqrt{2}R\right)dR\\
&= 10\sqrt{2} \int_{-\infty}^{\infty} r\left(R\right) \Phi(R)^3 \phi\left(\sqrt{2}R\right)dR + \frac{5}{\sqrt{2}}\left[l_3 - 3l_2 - l_1 \right]  + 12 l_2 - 6 l_1 \\
&= \frac{5}{\sqrt{2}}\left[l_3 - 3l_2 - l_1 \right]  + 12 l_2 - 6 l_1 
+ \frac{30}{8\pi\sqrt{3\pi}} \text{Arctan}\left(\frac{1}{4}\right) +
O\left(\int_{-\infty}^{\infty}R^9 \Phi(R) \exp\left(-R^2\right)dR\right)
\end{aligned}
\end{equation}

with 
\begin{equation}
\begin{split}
&I_1 = \frac{1}{2}\left(1 + \frac{1}{\sqrt{3}}\right),\\
&I_{2m+1} = \frac{(-1)^m m!}{2} - \frac{(-1)^m}{2}\left.\frac{\partial^m}{\partial p^m}\left(\frac{1}{p\sqrt{p+1/2}}\right)\right|_{p=1}\\
&\widetilde{I}_1 = \frac{1}{2}\left(1 - \frac{1}{\sqrt{3}}\right),\\
&G_{2n+1} = \frac{(-1)^n}{\sqrt{2}\pi}\left. \frac{\partial^n}{\partial p^n}\left(\frac{1}{p\sqrt{1/2 + p}}\text{Arctan}\left(\frac{1}{\sqrt{1 + 2p}}\right) \right)\right|_{p=1}\\
&\widetilde{G}_{2n+1} = \frac{(-1)^{n+1}}{\sqrt{2}\pi}\frac{\partial^n}{\partial p^n}\left(\frac{1}{p\sqrt{1/2 + p}}\text{Arctan}\left(\frac{-1}{\sqrt{1 + 2p}}\right) \right)
\end{split}
\end{equation}

\end{proof}

\subsection{Proof of Proposition \ref{PropGH}}

\begin{proof}
The first four population L-Moments of the g-and-h family of Tukey transform Loss models is given for $a=0$, $b=1$ by considering $$r(R) = \frac{\exp(gR)-1}{g}\exp\left(\frac{hR^2}{2}\right) $$  giving:
\begin{equation}
\begin{aligned}
l_{1} &=  \sqrt{2} \int_{-\infty}^{\infty} r\left(R\right) \phi\left(\sqrt{2}R\right)dR\\
&= \frac{\exp\left(\frac{g^2}{4 - 2 h}\right)\sqrt{2}}{g\sqrt{2 - h}} - \frac{2 }{g\sqrt{4 - 2 h}},\;\;\;\; h < 2\\
\end{aligned}
\end{equation}

\begin{equation}
\begin{aligned}
l_{2} &=  \sqrt{2} \int_{-\infty}^{\infty} r\left(R\right)\Phi(R) \phi\left(\sqrt{2}R\right)dR\\
&=  \frac{1}{g\sqrt{\pi}} \int_{-\infty}^{\infty}\Phi(R)\left[ \exp\left(-(1-h/2)R^2 + gR\right) - \exp\left(-R^2\right) \right] dR\\
&=  \frac{1}{2g\sqrt{\pi}}\int_{0}^{\infty}\left[1 + \text{erf}\left(\frac{R}{\sqrt{2}}\right)\right]\exp\left(-(1-h/2)R^2 + gR\right) dR\\
& + \frac{1}{2g\sqrt{\pi}} \int_{0}^{\infty}\left[1 + \text{erf}\left(\frac{R}{\sqrt{2}}\right)\right]\exp\left(-(1-h/2)R^2 - gR\right) dR - \frac{1}{g\sqrt{\pi}} \frac{\sqrt{2\pi}}{2}\\
&=  \frac{1}{2g\sqrt{\pi}} \frac{\exp\left(\frac{g^2}{(4 - 2 h)}\right) \sqrt{2 \pi}}{\sqrt{2 - h}} + \frac{1}{g\pi\sqrt{2}(1-h/2)}\psi_1\left(1, \frac{1}{2}, \frac{3}{2}, \frac{1}{2}; - \frac{1}{2(1-h/2)}, \frac{g^2}{4(1-h/2)} \right) - \frac{1}{g\sqrt{\pi}} \frac{\sqrt{2\pi}}{2}\\
\end{aligned}
\end{equation}

\begin{equation}
\begin{aligned}
l_{3} &=  \sqrt{2} \int_{-\infty}^{\infty} r\left(R\right) \left[3\Phi(R)^2 + 3\Phi(R) + 1\right] \phi\left(\sqrt{2}R\right)dR \\
&= 3l_2 + l_1 + \sum_{n=0}^{\infty} \frac{6(g)^{n-1}}{\sqrt{\pi}n!}\left\{
\frac{1}{4} 2^n (2 - h)^{-1 - n} \Gamma(1 + n) \right. \\
&\left. + \frac{2^{1 + n} (2 - h)^{-(3/2) - n}\Gamma(3/2 + n) {}_2F_1\left[1/2, 3/2 + n, 3/2, 
  1/(-2 + h)\right]}{2\sqrt{\pi}}\right. \\
& \left. + 1/4 \frac{(-1)^n 1}{\pi\sqrt{2}}\left.\frac{\partial^n}{\partial p^n}\left(\frac{1}{p\sqrt{1/2 + p}}\text{Arctan}\left(\frac{1}{\sqrt{1 + 2p}}\right) \right)\right|_{p=(1-h/2)} \right\}\\
\end{aligned}
\end{equation}

\begin{equation}
\begin{aligned}
l_{4} &=  \sqrt{2} \int_{-\infty}^{\infty} r\left(R\right) \left[10\Phi(R)^3 + 15\Phi(R)^2 + 12\Phi(R) - 6\right] \phi\left(\sqrt{2}R\right)dR\\
&=  12 l_2 - 6 l_1 + \frac{1}{g\sqrt{\pi}} 15 \sum_{n=0}^{\infty} \frac{(g)^{2n+1}}{(2n+1)!} \frac{(-1)^n }{\sqrt{2}\pi}\left.\frac{\partial^n}{\partial p^n}\left(\frac{1}{p\sqrt{1/2 + p}}\text{Arctan}\left(\frac{1}{\sqrt{1 + 2p}}\right) \right)\right|_{p=(1-h/2)}\\
&+\frac{45}{2g\sqrt{\pi}}\sum_{n=0}^{\infty} \frac{(g)^{2n+1}}{(2n+1)!} \frac{2^{1 + n} (2 - h)^{-(3/2) - n}\Gamma(3/2 + n) {}_2F_1\left[1/2, 3/2 + n, 3/2, 1/(-2 + h)\right]}{\sqrt{\pi}}\\
&+\frac{5}{g\sqrt{\pi}} \sum_{n=1}^{\infty} \frac{(g)^{n}}{n!} 2^{1/2 (-1 + n)} (1 + \exp(n \pi)) (2 - h)^{-(1/2) - n/2}\Gamma[(1 + n)/2]\\
&+ \frac{5}{4g\sqrt{\pi}} \sum_{n=1}^{\infty} \frac{(g)^{n}}{n!} \frac{1}{\pi (1-h/2)}\frac{3}{\sqrt{1 + 2(1-h/2)}}\text{Arctan}\left(\frac{1}{2\triangle\sqrt{1/2 + (1-h/2)}}\right),\\
& + O\left( \int_{-\infty}^{\infty} R^n \erf(R/\sqrt{2})^3 \exp\left(-(1-h/2)R^2\right)dR\right)
\end{aligned}
\end{equation}
with $\triangle = \sqrt{3/2 + (1-h/2)}.$
\end{proof}

\end{document}